\documentclass[aps,prl,reprint,amsmath,amssymb,superscriptaddress,floatfix]{revtex4-2}
\usepackage{amsthm,dsfont,ctable,url,braket,mathtools}
\usepackage{natbib}

\usepackage[colorlinks,
            linkcolor=blue,
            anchorcolor=blue,
            citecolor=blue,
urlcolor=blue
            ]{hyperref}

\usepackage{graphicx}
\usepackage[FIGTOPCAP,raggedright,nooneline]{subfigure}
\usepackage{float}
\usepackage{longtable}
\usepackage[T1]{fontenc}
\usepackage{lmodern}

\begin{document}
\title{Quantum verification of NP problems with single photons and linear optics}

\author{Aonan \surname{Zhang}}
\affiliation{National Laboratory of Solid State Microstructures, Key Laboratory of Intelligent Optical Sensing and Manipulation (Ministry of Education) and College of Engineering and Applied Sciences, Nanjing University, Nanjing 210093, China}
\affiliation{Collaborative Innovation Center of Advanced Microstructures, Nanjing University, Nanjing 210093, China}

\author{Hao \surname{Zhan}}
\affiliation{National Laboratory of Solid State Microstructures, Key Laboratory of Intelligent Optical Sensing and Manipulation (Ministry of Education) and College of Engineering and Applied Sciences, Nanjing University, Nanjing 210093, China}
\affiliation{Collaborative Innovation Center of Advanced Microstructures, Nanjing University, Nanjing 210093, China}

\author{Junjie \surname{Liao}}
\affiliation{National Laboratory of Solid State Microstructures, Key Laboratory of Intelligent Optical Sensing and Manipulation (Ministry of Education) and College of Engineering and Applied Sciences, Nanjing University, Nanjing 210093, China}
\affiliation{Collaborative Innovation Center of Advanced Microstructures, Nanjing University, Nanjing 210093, China}

\author{Kaimin \surname{Zheng}}
\affiliation{National Laboratory of Solid State Microstructures, Key Laboratory of Intelligent Optical Sensing and Manipulation (Ministry of Education) and College of Engineering and Applied Sciences, Nanjing University, Nanjing 210093, China}
\affiliation{Collaborative Innovation Center of Advanced Microstructures, Nanjing University, Nanjing 210093, China}

\author{Tao \surname{Jiang}}
\affiliation{National Laboratory of Solid State Microstructures, Key Laboratory of Intelligent Optical Sensing and Manipulation (Ministry of Education) and College of Engineering and Applied Sciences, Nanjing University, Nanjing 210093, China}
\affiliation{Collaborative Innovation Center of Advanced Microstructures, Nanjing University, Nanjing 210093, China}

\author{Minghao \surname{Mi}}
\affiliation{National Laboratory of Solid State Microstructures, Key Laboratory of Intelligent Optical Sensing and Manipulation (Ministry of Education) and College of Engineering and Applied Sciences, Nanjing University, Nanjing 210093, China}
\affiliation{Collaborative Innovation Center of Advanced Microstructures, Nanjing University, Nanjing 210093, China}

\author{Penghui \surname{Yao}}
\email{pyao@nju.edu.cn}
\affiliation{State Key Laboratory for Novel Software Technology, Nanjing University, Nanjing 210093, China}

\author{Lijian \surname{Zhang}}
\email[]{lijian.zhang@nju.edu.cn}
\affiliation{National Laboratory of Solid State Microstructures, Key Laboratory of Intelligent Optical Sensing and Manipulation (Ministry of Education) and College of Engineering and Applied Sciences, Nanjing University, Nanjing 210093, China}
\affiliation{Collaborative Innovation Center of Advanced Microstructures, Nanjing University, Nanjing 210093, China}

\date{\today}

\begin{abstract}
Quantum computing is seeking to realize hardware-optimized algorithms for application-related computational tasks. NP (nondeterministic-polynomial-time) is a complexity class containing many important but intractable problems like the satisfiability of potentially conflict constraints (SAT). According to the well-founded exponential time hypothesis, verifying an SAT instance of size $n$ requires generally the complete solution in an $O(n)$-bit proof. In contrast, quantum verification algorithms, which encode the solution into quantum bits rather than classical bit strings, can perform the verification task with quadratically reduced information about the solution in $\tilde{O}(\sqrt{n})$ qubits. Here we realize the quantum verification machine of SAT with single photons and linear optics. By using tunable optical setups, we efficiently verify satisfiable and unsatisfiable SAT instances and achieve a clear completeness-soundness gap even in the presence of experimental imperfections. The protocol requires only unentangled photons, linear operations on multiple modes and at most two-photon joint measurements. These features make the protocol suitable for photonic realization and scalable to large problem sizes. Our results open an essentially new route towards quantum advantages and extend the computational capability of optical quantum computing.
\end{abstract}

\maketitle 
\par\noindent
\textbf{Introduction}
\par 
Quantum computing has been found to unprecedentedly speed-up classically intractable computational tasks~\cite{Shor1997,10.1145/237814.237866,10.1145/1993636.1993682,Harrow2017,Preskill2018quantumcomputingin,
Arute2019,Zhongeabe8770}. As building universal, error-corrected quantum computers is still challenging, the community now seeks practical uses of noisy intermediate-scale quantum (NISQ) technologies in computational problems of interest and importance~\cite{Preskill2018quantumcomputingin}. Photonics has been a versatile tool in quantum information tasks~\cite{OBrien2009,AspuruGuzik2012,Flamini_2018} such as boson sampling~\cite{Broome794,Spring798,Tillmann2013,Crespi2013,Zhongeabe8770}, quantum walk~\cite{AspuruGuzik2012,Schreiber2012,Tang2018} and variational quantum simulation~\cite{Peruzzo2014,Santagatieaap9646}. By utilizing multi-degrees of freedom of photons~\cite{Kagalwala2017,PhysRevLett.120.260502} and well-developed linear optics~\cite{PhysRevLett.73.58,Carolan2015,Clements2016,PhysRevLett.124.010501}, information can be encoded and processed in a high-dimensional Hilbert space. These features make photonics a suitable platform to realize quantum algorithms involving high-dimensional encoding, low degree of entanglement and linear operations. Here we exploit
the advantages of photonics to realize a new regime of quantum algorithm --- the quantum verification machine (QVM) of nondeterministic polynomial-time (NP) problems. 
\par
The complexity class NP, which is the set of decision problems verifiable in polynomial time by a deterministic Turing machine, encompasses many natural decision and optimization problems. By definition, NP can be abstracted as a proof system which models computation as exchange of messages between the prover and the verifier. Verifying the correctness of a proof is a foundational computational model underpinning both the complexity theory and applications such as delegated computation. Specifically, we focus on the verification of the first discovered and most extensively studied NP-complete problem --- the Boolean satisfiability problem (SAT)~\cite{10.1145/800157.805047}, that is, the problem of asking whether a given Boolean formula with $n$ variables has a satisfying assignment. The NP-completeness signifies that any NP problem can be efficiently reduced to this problem. Corresponding to the problem of satisfying potentially conflict constraints, SAT has found numerous applications in circuit design, mode checking, automated proving and artificial intelligence~\cite{Malik2009}. Under the widely believed exponential time hypothesis (ETH)~\cite{Impagliazzo2001}, which asserts that the best algorithm for solving 3-SAT (a representative form of SAT) runs in time $2^{\gamma n}$ for some constant $\gamma>0$, verifying 3-SAT requires at least $O(n)$ bits. Otherwise the verifier can simply enumerate over all possible proofs which yield a sub-exponential algorithm for solving 3-SAT. Surprisingly, this bound on proof length no longer applies if quantum bits are used in proofs and verified by quantum computers. This perception rapidly aroused substantial efforts on quantum verification of NP(-complete) problems~\cite{v005a001,Blier2009,Chen2010,Chiesa2013,Harrow:2013:TPS:2432622.2432625,Brandao2017,Arrazola2018,Centrone2020}. In this line, Aaronson et al. proposed a protocol of proving 3-SAT with $O(\sqrt{n})$ unentangled quantum states each of $O(\log n)$ qubits~\cite{v005a001} and variants of the protocol have also been developed~\cite{Chen2010,Harrow:2013:TPS:2432622.2432625}. However, to date a complete demonstration of quantum verification algorithm is still missing.
\begin{figure*}
  \centering
  \includegraphics[width=\linewidth]{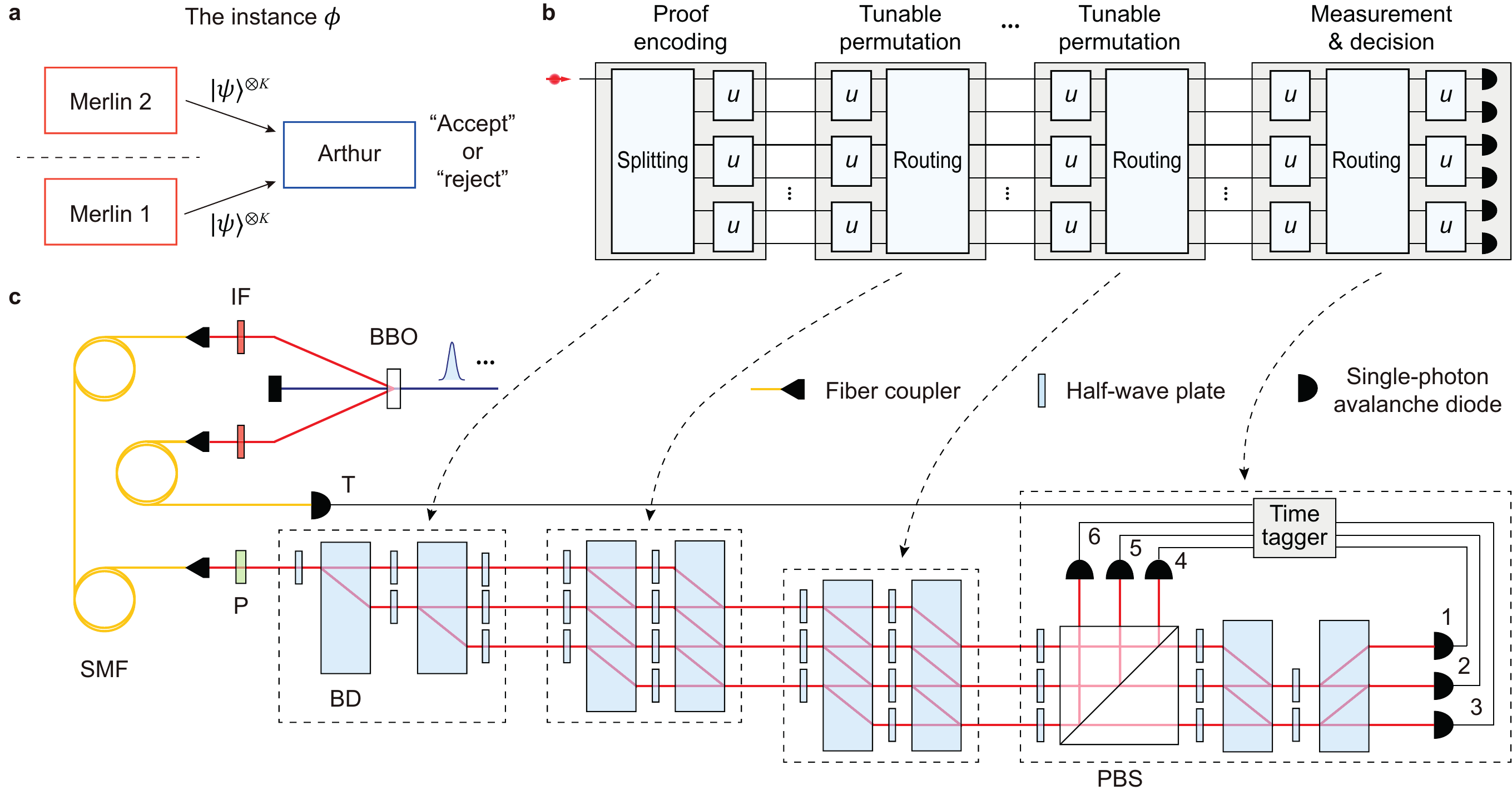}
\caption{\label{fig:setup} \textbf{Quantum verification machine.} \textbf{a}, The two-prover Quantum Merlin-Arthur protocol [QMA(2)]. On the basis of the given SAT instance, the two Merlins send unentangled, identical proof states to Arthur, who checks the proof on his quantum computer and makes an ``accept'' or ``reject'' decision. \textbf{b}, The architecture of quantum circuit for the satisfiability and uniformity test. The design comprises proof encoding, tunable permutations and measurement on the modes. These operations are mainly based on tunable two-mode transformations $u$ combined with mode splitting and routing. With the input of single photons, the circuit can verify the satisfiability of a set of clauses or the uniformity on random matchings. \textbf{c}, Experimental setup for the satisfiability test and uniformity test. Merlins prepare single photons distributed in the polarization and path modes, encoding the assignment in the single-photon states as quantum witnesses. Arthur then applies permutations and interferences on these modes with linear optics. Note only states from one Merlin are required for the two tests. The output modes are detected by single-photon avalanche diodes (SPADs) and registered by a time tagger. BBO, $\beta$-barium borate crystal; BD, calcite beam displacer; P, polarizer; IF, interference filter; SMF, single mode fibre; PBS, polarizing beam-splitter.}
\end{figure*}
\par
In this work, we report the first experimental quantum verification of SAT with single photons and linear optics, by implementing a modified version of recent proposals~\cite{Arrazola2018}. We present a scalable design of reconfigurable optical circuits in which quantum proofs are mapped to single photons distributed in optical modes. The experiment demonstrates faithful verification of NP problems in terms of a complete analysis on the satisfiable instance, unsatisfiable instance and cheating prover cases. Our work links the remarkable proof systems in computer science to the manipulation and detection of photons, which foreshadows further investigations of a variety of computational models in the photonic regime.
\par\noindent
\textbf{Results}
\par\noindent
\textbf{Quantum verification algorithm of the satisfiability problem. }An instance of SAT is formalized as the conjunction of a set of clauses $\phi=c_1 \land c_2 ...\land c_j$, each of which is the disjunction of a set of literals $l_1 \lor l_2 ...\lor l_m$. A literal could be a variable $x_i$ or a negation of a variable $\neg {x_i}$. In 3-SAT instances, each clause has exactly 3 literals. The quantum verification of 3-SAT corresponds to the complexity class Quantum Merlin-Arthur [QMA($K$)], as the quantum analogue of NP~\cite{10.5555/863284,892141,10.1007/978-3-540-24587-2_21}. In this scheme, $K$ non-communicating, omniscient provers (called Merlins) send $K$ unentangled quantum proofs to a skeptical, computationally bounded verifier Arthur to convince Arthur the instance is satisfiable (see Fig. \ref{fig:setup}a). Arthur checks the proof in his computing machines and decide whether to accept or reject the proof. Two properties are required in a QMA protocol: (i) \textit{Completeness}: if the instance is satisfiable, there exist a proof such that Arthur accepts with at least some high probability $c$; (ii) \textit{Soundness}: if the instance is not satisfiable, for any proof Arthur accepts with at most some probability $s$. 
\par
The protocol firstly reduces the 3-SAT instance to a 2-out-of-4 SAT instance where each clause contains four variables $x_i,x_j,x_k,x_l$ and is satisfied if two of them are true, i.e. $x_i+x_j+x_k+x_l=2$. In the verification, Merlins are supposed to send Arthur $K=O(\sqrt{n})$ identical, unentangled quantum states~\cite{v005a001}, each of the form
\begin{equation}
\label{eq:proof}
|\psi\rangle=\frac{1}{\sqrt{n}}\sum_{i=1}^n(-1)^{x_i} |i\rangle,
\end{equation}
where $|i\rangle=\hat{a}^\dagger_i |0\rangle$ and $\hat{a}^\dagger_i$ is the creation operator on mode $i$. Here $x_1, x_2,..., x_n \in \{0,1\}^n$ is an assignment of the $n$ variables. A state of such form is called a proper state. The $n$-dimensional quantum state can be equivalently described by $\log n$ qubits revealing at most $\log n$ bits information by measurements on the state. To check whether the assignment $x$ satisfies the clauses, Arthur can choose some clauses $(i,j,k,l)$ at random and measure the $K$ copies of $|\psi\rangle$ in a basis with a projection on $|c\rangle=(|i\rangle+|j\rangle+|k\rangle+|l\rangle)/2$ for each clause. For each copy Arthur will get a probability of observing the outcome $|c\rangle$
\begin{equation}
p_c=|\langle c|\psi\rangle|^2=[(-1)^{x_i}+(-1)^{x_j}+(-1)^{x_k}+(-1)^{x_l}]^2/4n. \nonumber
\end{equation}
Then Arthur rejects the proof if he gets the outcome $|c\rangle$ for at least one copy and accepts it otherwise. With this \textit{Satisfiability Test}, Arthur will have $p_c=0$ if $x_i+x_j+x_k+x_l=2$, and some constant non-zero probability otherwise. An issue is that Merlins may cheat Arthur by sending him improper state, for example concentrating the amplitude in a subset of the basis $\{|i\rangle\}$ such that the \textit{Satisfiability Test} passes even the instance is not satisfiable. To tackle this problem Arthur can perform \textit{Uniformity Test}: he randomly chooses a matching $M$ on the set $\{1,...,n\}$ such that the set is partitioned into $n/2$ groups of the form $(i,j)$, then measures each copy of the state $|\psi\rangle$ in the basis with $\{|i\rangle+|j\rangle,|i\rangle-|j\rangle\}$ for each $(i,j)\in M$. Only if the state is proper (i.e. the amplitudes are equal), one of the two outcomes will never occur. With the statistics on the outcomes, Arthur rejects the proof if two outcomes $\{|i\rangle+|j\rangle,|i\rangle-|j\rangle\}$ both occur for a same $(i,j)\in M$. Here the $K$ copies are used to obtain sufficient statistics on the outcomes to make a decision.
\par
As the verification requires multiple copies of the state, another possible way for Merlins to cheat is to send different states rather than identical copies. For this reason, Arthur performs \textit{Symmetry Test}: a swap test between two states, which accepts with certainty if the two states are identical and has a constant probability to reject when the two-state overlap is under a certain threshold. The QMA($K$) protocol may be significantly reduced by simulating the $K$ Merlins with a single Merlin who sends a product state of the $K$ copies $|\psi\rangle^{\otimes K}$, yet in this case Arthur needs to guarantee the unentanglement among the $K$ subsystems. To this end Arthur can ask for the proof state $|\psi\rangle^{\otimes K} \in {\mathbb{C}_d }^{\otimes K}$ from another Merlin and conduct a \textit{Product Test}~\cite{Harrow:2013:TPS:2432622.2432625}, which applies the swap test to each of the $K$ pairs of corresponding subsystems of the two states. The proof will be accepted if all the swap tests pass and rejected otherwise. With the help of the product test, we can simulate the $K$-prover protocol with only 2 Merlins, which corresponds to the complexity result QMA($K$)=QMA(2) for $K\geq 2$~\cite{Harrow:2013:TPS:2432622.2432625}.
\par
Overall, Arthur performs one of the four aforementioned tests with constant probability (e.g. 1/4 each). As a consequence, we have an efficient quantum algorithm to verify SAT with perfect completeness and constant soundness, using two unentangled proofs of length $O(\sqrt{n}\log n)$ qubits (see Materials and methods for a summary of the protocol).
\begin{figure*}
  \centering
  \includegraphics[width=\linewidth]{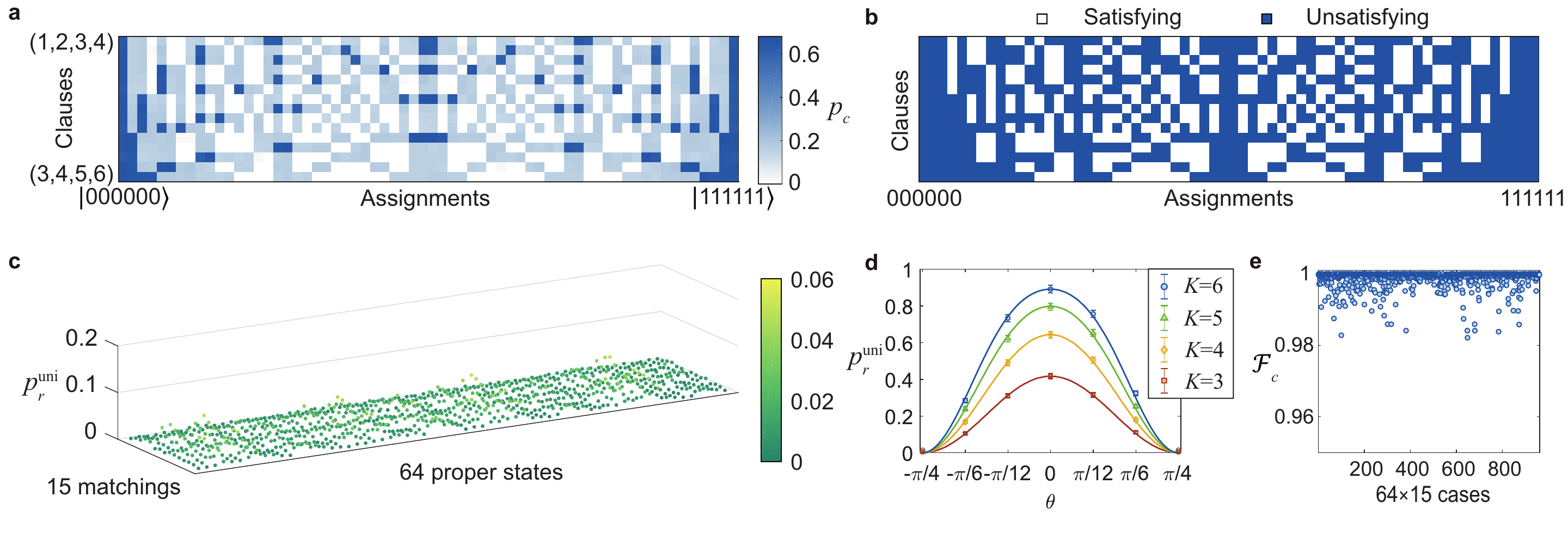}
\caption{\label{fig:satuni} \textbf{Validation of the satisfiablity test and uniformity test.} \textbf{a}, The experimentally measured projection probabilities $p_c$ when verifying the 15 clauses (rows) with the 64 proper states (columns). Here the proof states from left to right are $|000000\rangle,|100000\rangle...,|111111\rangle$, while the verified clauses from top to down are $(1,2,3,4),(1,2,3,5)...,(3,4,5,6)$. $(i,j,k,l)$ denotes the clause $x_i+x_j+x_k+x_l=2$. \textbf{b}, The satisfiability of the 15 clauses for the 64 assignments. An assignment $x$ may satisfies (white) or unsatisfies (blue) a certain clause. \textbf{c}, The measured rejection probabilities $p_r^\text{uni}$ of the uniformity test for the 64 proper states $\times$ 15 matchings when the number of copies $K=3$. \textbf{d}, The rejection probability $p_r^\text{uni}$ of the uniformity test for improper states of the form $|\psi_{\text{im}}(\theta)\rangle=(\cos \theta,\sin \theta,\cos \theta,\sin \theta,\cos \theta,\sin \theta)/\sqrt{3}$. Here we take the results for the matching $\{(1,2),(3,4),(5,6)\}$ under different numbers of copies as an example. For each $\theta$ we run the test 5000 times and collect the measurement outcomes of $5000 \times K$ photons to acquire the rejection probability $p_r^\text{uni}$. The results given by numerical simulations are shown as solid lines. Error bars are uncertainties assuming Poisson count statistics. \textbf{e}, The statistical fidelities $\mathcal{F}_c=\left(\sqrt{p_c^\text{the} p_c^\text{exp}}+\sqrt{(1-p_c^\text{the}) (1-p_c^\text{exp}})\right)^2$ between the theoretical probabilities $p_c^\text{the}$ and experimental probabilities $p_c^\text{exp}$ for the $64 \times 15$ cases of the satisfiability test.}
\end{figure*}
\par\noindent
\textbf{Photonic implementation of the quantum verification machine. }To realize the verification algorithm in photonic regime, we devise optical circuits for the four tests and experimentally implement the circuit in the case $n=6$. The proofs from the two Merlins are unentangled photons generated by a parametric down-conversion process while the $K$ copies of the state $|\psi\rangle$ correspond to photons generated sequentially at different time. In our experiment the $K$ copies sent by a same Merlin are identical due to the fact that the apparatus to prepare the states is fixed within the duration of the experiment. For each copy we encode the $n$-dimensional quantum state in the polarization and path degrees of freedom of the photon. The optical modes $\{|1\rangle,|2\rangle,|3\rangle,|4\rangle,...,|n\rangle\}$ are mapped to $\{|h_1\rangle,|v_1\rangle,|h_2\rangle,|v_2\rangle,...,|v_{n/2}\rangle\}$, where $|h_j\rangle$ ($|v_j\rangle$) denotes the horizontal (vertical) polarization in path $j$. In the following we use $|x_1 x_2 x_3 x_4 x_5 x_6 \rangle$ to represent a proper state given in equation (\ref{eq:proof}) encoding the assignment $x_1 x_2 x_3 x_4 x_5 x_6$. When $x_i=0$ the phase on mode $i$ is 0, whereas $x_i=1$ the phase is $\pi$.
\par
Figure \ref{fig:setup}b depicts the circuit design for the satisfiability test and uniformity test. The circuit comprises a sequence of stages, each of which involves a set of two-mode configurable transformations $u$ combined with mode splitting or routing (see Materials and methods for details). Starting from proof encoding, Merlin firstly splits the input single photon into an equal superposition over $n$ modes and encodes the assignment $x$ into the $K$ copies of the state. Each state is then sent to successive tunable permutation modules, which select the modes corresponding to the chosen clause $(i,j,k,l)$ or group the modes into a random matching $M$. Finally, the measurement and decision module performs either projection on the certain state $|c\rangle$ or two-mode interferences on the certain matching $M$. The two-mode transformations $u$ are implemented by half-wave plates (see Fig. \ref{fig:setup}c), of which the optical axes can be set in different angles to perform different two-mode sub-operations such as Pauli-$X$, Pauli-$Z$ and Hadamard gates
\begin{equation}
X=\frac{1}{\sqrt{n}}\begin{pmatrix}
0 & 1  \\
1 & 0 
\end{pmatrix},
Z=\frac{1}{\sqrt{n}}\begin{pmatrix}
1  & 0  \\
0 & -1 
\end{pmatrix},
H=\frac{1}{\sqrt{2n}}\begin{pmatrix}
1  & 1  \\
1 & -1 
\end{pmatrix}. \nonumber
\end{equation}
With appropriate configurations of these gates, the circuit can perform different permutations and interferences on the optical modes. The ability of the permutation stage is to sort the modes into groups (2 or 4 modes each, without regard to order). Configurations of the optical circuit are designed to realize the $\dbinom{6}{4}=15$ projections and the $\dbinom{6}{2}\times\dbinom{4}{2}\div 3!=15$ matchings. The measurement outcome is read out by single-photon avalanche diodes and we register the measurement outcome for each copy of the proof state with a multi-channel time tagger. For a single trial of the test, a decision on the proof (``reject'' or ``accept'') is made based on the detector pattern of $K$ copies: for the satisfiability test, whether the detector corresponding to the projector $|c\rangle\langle c|$ clicks; for the uniformity test, whether the two detectors in a same group $(i,j)$ both click.
\par\noindent
\textbf{Quantum verification of SAT instances with linear optics. }Firstly we demonstrate the performance of the verifier in the satisfiability and uniformity tests. By changing the settings of the wave plates to prepare the 64 proper states and verify the 15 clauses, we measure the probabilities $p_c$ for all the $64 \times 15$ cases (Fig. \ref{fig:satuni}a), which are consistent with the theoretical satisfiability of the clauses (Fig. \ref{fig:satuni}b). The satisfying proofs manifest nearly zero outcome probabilities ($0.28\%$ in average), whereas all the unsatisfying proofs manifest significant outcome probabilities exceeding the probabilities of the satisfying cases by two orders of magnitude (larger than $13.47\%$). Regarding the uniformity test, we show the rejection probabilities when testing the 64 proper states for the 15 matchings with $K=3$ in Fig. \ref{fig:satuni}c. The results exhibit a high probability of $98.67\%$ to accept in average. For the case that Merlins send improper states, we run the uniformity test for proof states of the form $|\psi_{\text{im}}(\theta)\rangle=(\cos \theta,\sin \theta,\cos \theta,\sin \theta,\cos \theta,\sin \theta)/\sqrt{3}$ with different numbers of copies $K=3,4,5,6$ (Fig. \ref{fig:satuni}d). Here $(\alpha_1,\alpha_2,\alpha_3,\alpha_4,\alpha_5,\alpha_6)$ denotes a state with complex amplitudes $\alpha_i$ in mode $|i\rangle$, i.e. $\sum_{i=1}^n \alpha_i |i\rangle$. An increase in the rejection probability is observed with the transition from proper states to highly improper states, which fits the numerical simulations. On the other hand, higher rejection probabilities are obtained for improper states when increasing the number of copies $K$. In addition, we determine the average statistical fidelity $\mathcal{F}_c=\left(\sqrt{p_c^\text{the} p_c^\text{exp}}+\sqrt{(1-p_c^\text{the}) (1-p_c^\text{exp}})\right)^2$ between the theoretical and experimental projection probabilities ($p_c^\text{the}$ and $p_c^\text{exp}$) to be $0.9988 \pm 0.0024$ (see Fig. \ref{fig:satuni}e), which justifies the excellent agreements between experimental results and theoretical calculations.
\begin{figure*}
  \centering
  \includegraphics[width=0.95\linewidth]{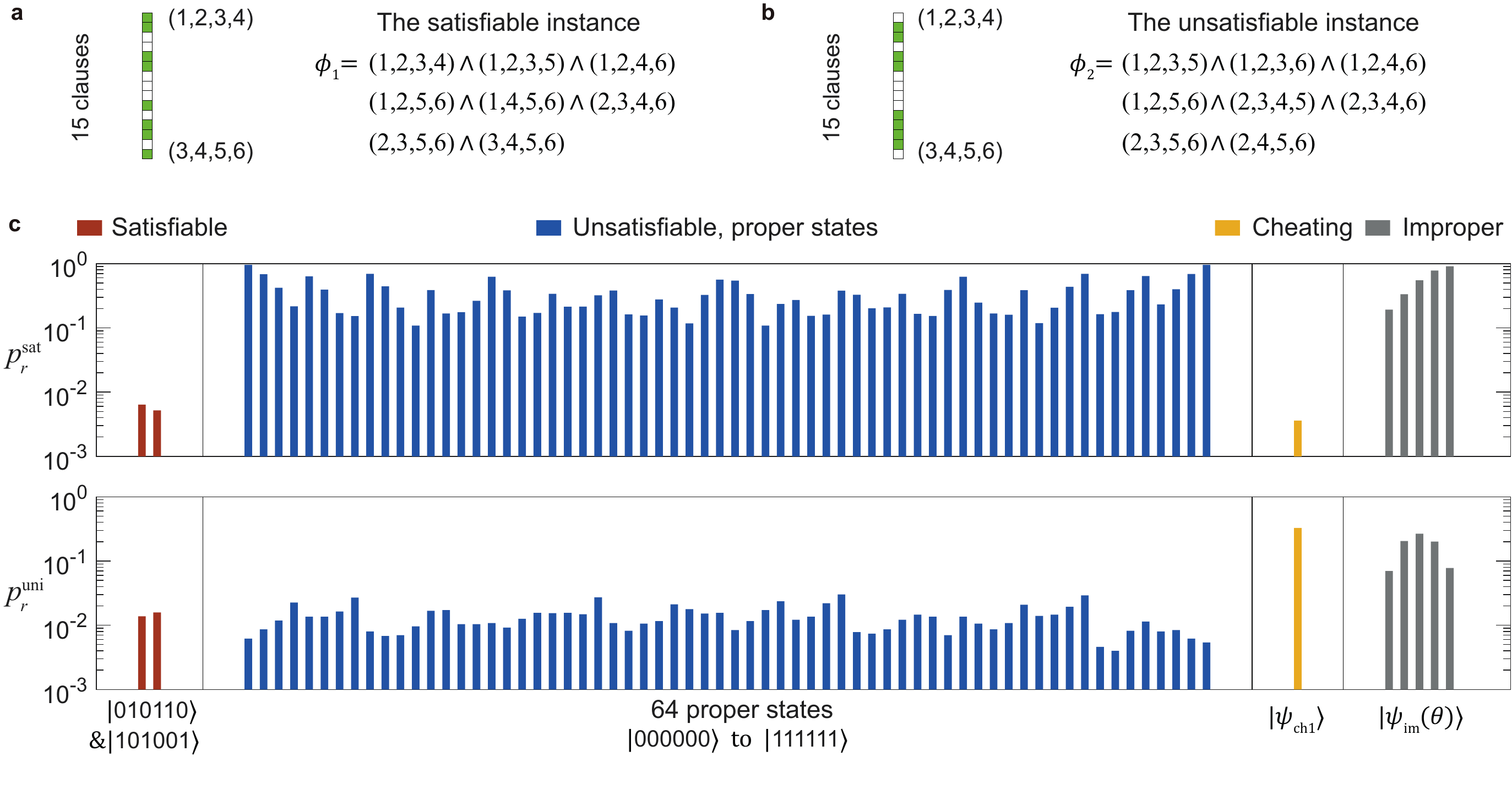}
\caption{\label{fig:example} \textbf{Experimental verification of SAT instances.} \textbf{a}, The satisfiable instance $\phi_1$. \textbf{b}, The unsatisfiable instance $\phi_2$. The shaded squares (green) illustrate which 8 of the 15 clauses are chosen in the instance. \textbf{c}, The rejection probabilities of the satisfiability test ($p_r^\text{sat}$, top) and uniformity test ($p_r^\text{uni}$, down) for different proof states. For the satisfiable instance $\phi_1$, Merlins will send proof states encoding the correct solution thus we show the results for the two satisfying proof states (red bars). For the unsatisfiable instance $\phi_2$, we test different cases consisting of sending the 64 proper states (blue bars), a deliberate cheating state $|\psi_\text{ch1}\rangle$ in order to pass the satisfiability test (yellow bars), and improper states $|\psi_{\text{im}}(\theta)\rangle$ (grey bars). The number of copies $K=3$ is adopted in the verification and the rejection probabilities are obtained by repeating each test 5000 times.}
\end{figure*}
\par
To demonstrate the verification of specific instances, we concentrate on the instances including 8 clauses, in which there are $\dbinom{15}{8}=6435$ instances. According to the satisfiability of the clauses (Fig. \ref{fig:satuni}b), 90 instances are satisfiable (each with two solutions) and 6345 instances are unsatisfiable. Figure \ref{fig:example} visualizes the results of verifying a satisfiable instance $\phi_1$ (illustrated in Fig. \ref{fig:example}a) and an unsatisfiable instance $\phi_2$ (illustrated in Fig. \ref{fig:example}b). As Merlins aim to make Arthur accept the proof, for the satisfiable instance $\phi_1$ Merlins will honestly send the proof encoding one of the two satisfying assignments. In this case the proof states successfully pass both tests with high probabilities ($p_r^\text{sat}=0.64\%$ and $p_r^\text{uni}=1.31\%$, averaging over the two states), as shown in Fig. \ref{fig:example}c. 
\par
For the unsatisfiable instance $\phi_2$, we consider situations where Merlins send different types of states (Fig. \ref{fig:example}c). Firstly we perform the two tests with all the 64 proper states. The verifier attains rejection probabilities $p_r^\text{sat}$ larger than $11.50\%$ and up to $95.72\%$ in the satisfiability test although these proofs could probably pass the uniformity test ($p_r^\text{uni}=1.30\%$ averaging over the 64 proper states). Secondly we realize cheating Merlins by sending deliberately designed improper states in order to pass the satisfiability test. As an example, we construct the state $|\psi_{\text{ch1}}\rangle=(1,-3,1,1,1,1)/\sqrt{14}$ (as well as $|\psi_{\text{ch2}}\rangle=(-3,1,1,1,1,1)/\sqrt{14}$ for instances given in the Supplementary Information), for which the projection probability $p_c$ of verifying any of the eight clauses in $\phi_2$ theoretically equals zero. Consequently, $|\psi_{\text{ch1}}\rangle$ reaches a rejection probability $p_r^\text{sat}=0.44\%$ of the same order of magnitude as in the satisfiable case. Nevertheless, Arthur can detect the cheating with the help of the uniformity test, in which a rejection probability of $31.90\%$ is obtained. This result justifies the necessity of the uniformity test. Finally the verification is also executed by sending just improper states $|\psi_{\text{im}}(\theta)\rangle$ with $\theta=\{-\pi/6,-\pi/12,0,\pi/12,\pi/6\}$, which exhibit considerable rejection probabilities in both tests. We conclude from the results that for all the three cases, evident rejection probabilities are observed in at least one of the two tests. The typical realizations indicate close to perfect completeness and constant soundness and thereby experimentally achieve a clear completeness-soundness gap for the quantum verification (see Supplementary Information for more examples and results). Experimental imperfections, including the limited interference visibilities, phase fluctuations and errors in the operations, lead to deviations of the outcome probabilities from ideal ones for the satisfying proof states and thereby imperfect completeness for the protocol. In real-world applications of the QVM, of particular importance is the amplification of the completeness-soundness gap. For this reason we also demonstrate the amplification of the success probability for the instances $\phi_1$ and $\phi_2$, of which the protocol and results are given in the Supplementary Information.
\begin{figure*}
  \centering
  \includegraphics[width=\linewidth]{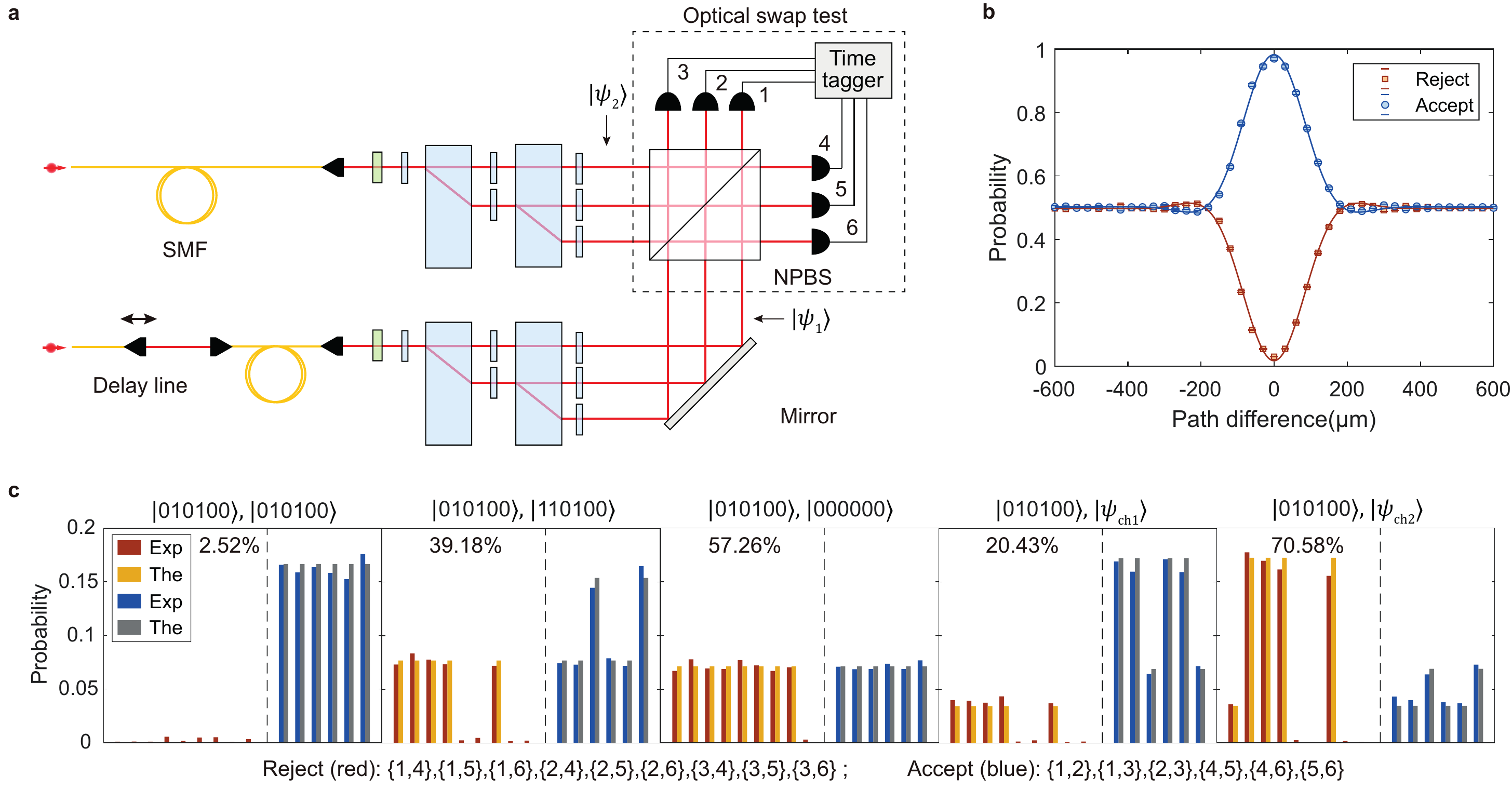}
\caption{\label{fig:hom} \textbf{The optical swap test.} \textbf{a}, Experimental scheme. Two single photons are injected into the setup and prepared as two quantum proofs $|\psi_1\rangle$ and $|\psi_2\rangle$. The two states $|\psi_1\rangle$ and $|\psi_2\rangle$ are interfered at a non-polarizing beam-splitter (NPBS), and the interference results are read out by detectors. A time tagger registers single-shot events from all the two-fold coincidence channels. The path difference between the two photons can be changed by a delay line to observe the interference. \textbf{b}, Multi-dimensional Hong-Ou-Mandel (HOM) interference. Solid lines are curve fittings of the data to a Gaussian multiplied by a sinc function. A HOM interference dip (peak) is observed for the rejection (corrected acceptance) probabilities. Error bars are uncertainties assuming Poisson count statistics. \textbf{c}, The results of the swap test for typical cases: the two states are proper and the same (the first panel); the two states are proper but not the same (the second and third panels), one of the state is proper and the another is improper (the fourth and fifth panels). Each panel shows the experimental (red and blue bars) and theoretical (yellow and grey bars) outcome probabilities on the 15 coincidence channels. The percentage labelled in each panel denotes the rejection probability of the swap test $p_r^{\text{swap}}$.}
\end{figure*}

The symmetry test and the product test require optical swap test~\cite{PhysRevA.87.052330}, which can be implemented with a multi-mode Hong-Ou-Mandel (HOM) interference (Fig. \ref{fig:hom}a)~\cite{PhysRevLett.59.2044}. Our experiment uses a non-polarizing beam-splitter (NPBS) to perform the two-photon interferences on the six optical modes distributed in both polarization and path degrees of freedom. In the optical swap test, the probability of rejection is $p_r^{\text{swap}}=(1-|\langle \psi_1|\psi_2\rangle|^2)/2$, where $|\psi_1\rangle$ and $|\psi_2\rangle$ are the photonic states in the two input ports of the NPBS. We register all the $\dbinom{6}{2}=15$ coincidence channels, in which the 6 one-side channels (the two photons are detected in the same output port of the NPBS) correspond to the ``accept'' outcome and the 9 two-side channels (the two photons are detected in different output ports of the NPBS) correspond to the ``reject'' outcome. We change the path difference between the two states with a delay line and observe the high-dimensional two-photon HOM interference. The HOM interference of identical proper states (Fig. \ref{fig:hom}b) manifests peaks for the ``accept'' outcomes and dips for the ``reject'' outcomes, resulting in a high acceptance probability of $(97.48\pm 0.56)\%$. This result guarantees a high probability to accept in product test, as an experimental demonstration of the reduction from QMA($K$) to QMA(2). To demonstrate the performance of the symmetry test, we apply the optical swap test to different combinations of states, as shown in Fig. \ref{fig:hom}c. On the basis of the outcome probabilities over the detector patterns, it can be concluded that considerable probabilities are obtained in the ``reject'' outcomes if the two states are not the same. The theoretical predictions also agree with the experimental results.
\par\noindent
\textbf{Discussion}
\par\noindent
The results of the four tests, which constitute a complete quantum verification of SAT, highlight the capability of photonic machines to realize a new type of quantum advantage on the computational space~\cite{Maslov2020}. Through the lens of computational complexity, the quantum provers reveal $O(\sqrt{n}\log n)$-bit information, whereas classical provers in the best algorithm need to reveal $O(n)$ bits, not better than simply writing down the complete solution. The quantum verification machines driven by $\tilde{O}(\sqrt{n})$ qubits can efficiently carry out the classically impossible computation, breaking through the $O(n)$-bit limit for classical algorithms imposed by ETH. If we in turn focus on the task of NP verification with limited information, a classical computer with an $O(\sqrt{n}\log n)$-bit message runs in exponential time $2^{O(n-\sqrt{n}\log n)}$ just assuming ETH, whereas the quantum algorithm runs in a polynomial-time overhead~\cite{Arrazola2018}. Consequently, QVMs will show an exponential speed-up over classical computers with limited information. Developments on quantum computation pursue provable quantum-classical separation. As ETH is a well-founded complexity-theoretic conjecture in computer science, our result foreshadows a desirable route towards realizing quantum advantages in an useful problem under a ``fine-grained'' complexity assumption~\cite{Harrow2017}.
\par
We have demonstrated the quantum verification algorithm of the satisfiability problem with two unentangled quantum witnesses, using single photons and tunable optical circuits. By combining algorithmic designs and experimental realizations, we optimize the whole architecture of the optical circuit and realize faithful verification of instances with high accuracies and scalability. Our demonstration extends the capability of optical quantum computing into the significant computational model of proof verification. Scaling up the scheme, which requires large scale programmable linear-optical systems and precise control of experimental imperfections, is an appealing route towards quantum advantage. With current advances in photonic technologies~\cite{OBrien2009,AspuruGuzik2012,Flamini_2018,Wang2019}, we expect this scheme can be scaled to larger problem sizes in the near future. Among substantial prospects, we envision quantum verification machines can stimulate experimental studies of various proof systems (QMA, QAM, QIP, MIP* etc~\cite{10.5555/863284,892141,10.1007/978-3-540-24587-2_21,10.1145/2049697.2049704,Ji2020}), inspire future developments of verifier-based quantum algorithms, and find applications in cloud-based quantum computing~\cite{Barz2012,Reichardt2013,Barz2013,Fisher2014}. Our work opens a new avenue in the utility of photonic NISQ devices and adds a key ingredient to the investigation towards answering valuable questions on both computational complexity and quantum physics.

\par\noindent
\textbf{Materials and methods}
\par\noindent
\textbf{Quantum verification algorithm.} The class Quantum Merlin-Arthur [QMA($K$)] consists of the set of decision problems having $K$ unentangled polynomial-size quantum proofs that can be verified on a quantum computer in polynomial time. As the quantum analogue of the complexity class nondeterministic-polynomial-time (NP), QMA($K$) has received extensive interests and many natural problems are proven to be in the class, such as $N$-representability~\cite{PhysRevLett.98.110503} in quantum chemistry. Formally, a language $L$ is in QMA$(K)_{c,s}$ if there exists a polynomial-time quantum algorithm $V$ such that, for all inputs $x\in\{0,1\}^n$: 
\par
(i) \textit{Completeness}. If $x\in L$, there exists $K$ witnesses with poly($n$) qubits each, such that $V$ outputs ``accept'' with probability at least $c$.
\par
(ii) \textit{Soundness}. If $x\notin L$, $V$ outputs ``accept'' with probability at most $s$ for all proof states.
\par
Our quantum verification algorithm is a modified version of the recent proposals~\cite{v005a001,Harrow:2013:TPS:2432622.2432625,Arrazola2018}. The protocol proceeds as follows.
\par 
Given a 2-out-of-4 SAT instance $\phi$, each of the two Merlins sends to Arthur a quantum state in ${\mathbb{C}_n }^{\otimes K}$ (with $K$ subsystems). The two quantum states are denoted as $|\varphi_1\rangle$ and $|\varphi_2\rangle$ respectively. Arthur performs one of the following four tests, each with probability 1/4.
\par
(1) \textit{Satisfiability Test}. Arthur randomly chooses a block containing a set of clauses such that no variable appears more than once. Then Arthur measures each of the $K$ subsystems from Merlin 1 in a basis corresponding to the clauses in the block. For each clause $(i,j,k,l)$, Arthur performs the projection on $|c\rangle=(|i\rangle+|j\rangle+|k\rangle+|l\rangle)/2$. If the outcome $|c\rangle$ is obtained for at least one subsystem, reject. Otherwise, accept.
\par
(2) \textit{Uniformity Test}. Arthur randomly chooses a matching $M$ on the set $\{1,2,...,n\}$, and measures each of the $K$ subsystems from Merlin 1 in a basis containing $\{(|i\rangle+|j\rangle)/\sqrt{2},(|i\rangle-|j\rangle)/\sqrt{2}\}$ for every edge $(i,j)\in M$. If for some edge $(i,j)$, the two outcomes $(|i\rangle+|j\rangle)/\sqrt{2}$ and $(|i\rangle-|j\rangle)/\sqrt{2}$ both occur, reject. Otherwise, accept.
\par
(3) \textit{Symmetry Test}. Arthur chooses the subsystem 1 and another randomly chosen subsystem from Merlin 1, and performs a swap test on the two states. If the swap test passes, accept. Otherwise, reject.
\par
(4) \textit{Product Test}. Arthur performs swap test on each of the $K$ pairs of corresponding subsystems of $|\psi_1\rangle$ and $|\psi_2\rangle$, and accepts if all of the swap tests pass. Otherwise, reject.
\par
\textbf{Photon source. }Frequency-doubled light pulses ($\sim$150 fs duration, 415 nm central wavelength) originating from a Ti:Sapphire laser (76 MHz repetition rate; Coherent Mira-HP) pump a beta barium borate ($\beta$-BBO) crystal phase-matched for type-II beamlike spontaneous parametric downconversion (SPDC) to produce degenerate photon pairs (830 nm central wavelength). The photon pairs are spectrally filtered by interference filters (IF) with 3 nm full-width at half-maximum and collected into single mode fibres (SMF). The pump power is set to $\sim$150 mW to ensure a low probability of emitting two photon pairs. By detecting one of the pair via a single-photon avalanche diode, we characterize the second order correlation function of heralded single photons to be $g^{(2)}(0)=0.041\pm 0.008$. A Hong-Ou-Mandel interference visibility $\mathcal{V}=0.969\pm 0.004$ is observed, indicating a great indistinguishability between the two photons. The high indistinguishability guarantees a good performance of the optical swap test. See Supplementary Information for details about the $g^{(2)}(0)$ measurements and the HOM interference.
\par
\textbf{Optical circuit. } In the satisfiability test and uniformity test, Arthur merely requires to measure the quantum proof $|\psi\rangle^{\otimes K}$ from one Merlin (Merlin 1 in the experiments), therefore the optical circuit shown in Fig. \ref{fig:setup}\textbf{b} is designed to perform local operations with the input of a single photon in each measurement. The single photons generated in the SPDC source are firstly delivered to polarization controllers and polarizers to prepare horizontally polarized states and then directed towards the optical circuit. The circuit is divided into three stages: (i) proof encoding; (ii) a sequence of tunable permutations; (iii) measurement and decision.
\par
In Stage (i), firstly the input single photon passes the splitting module and evolves to an equal superposition on $n/2$ optical modes
\begin{equation}
\hat{a}^\dagger_1 |0\rangle \mapsto \sqrt{\frac{2}{n}}\left(\sum_{j=1}^{n/2-1} \hat{a}^\dagger_{2j-1}+\hat{a}^\dagger_{n}\right) |0\rangle.
\end{equation}
Here ${|0\rangle}$ denotes the vacuum state. This evolution is experimentally realized by a sequence of wave plates and calcite beam displacers. The following operation is a combination of $n/2$ two-mode transformations $\{u_j(\theta_j)\}$, which constitute an $n$-mode transformation
\begin{equation}
U=\bigoplus_{j=1}^{n/2}u_j(\theta_j).
\end{equation}
Each two-mode transformation $u_j(\theta_j)$ can be written as 
\begin{equation}
u_j(\theta_j)=\begin{pmatrix}
\cos \theta_j  & \sin \theta_j  \\
\sin \theta_j & -\cos \theta_j
\end{pmatrix} ,
\end{equation}
where the angle of the optical axis of the corresponding half-wave plate is $\theta_j/2$. Each wave plate can be configured into one of the four different angles to prepare equal superposition encoding the assignment of the two variables $(x_{2j-1},x_{2j})$ as 00,01,10 or 11. As a result, the overall transformation $U$ can prepare arbitrary proper states. For the cheating Merlins, the wave plates are set into angles differing from the honest case to implement an unequal splitting and (or) a different transformation $U$. The details on proof encoding are given in the Supplementary Information.
\par
Stage (ii) comprises a sequence of tunable permutations, each consisting of a transformation $U$ and a mode routing. In this case the two-mode transformations $\{u_j(\theta_j)\}$ are set to two-mode $X$ or $Z$ operations to permutate the two modes or not. The operation of the mode routing is equivalent to a fixed permutation. For example, one of the permutation matrix for mode routing in our experiment can be described as
\begin{equation}
P_0=\begin{pmatrix}
 0& 1& 0& 0& 0& 0 \\
 0& 0& 0& 1& 0& 0 \\
 1& 0& 0& 0& 0& 0 \\
 0& 0& 0& 0& 0& 1 \\
 0& 0& 1& 0& 0& 0 \\
 0& 0& 0& 0& 1& 0
 \end{pmatrix}. \nonumber
\end{equation}
The combination of the aforementioned two operations enables programmable permutation $P\cdot U$ on the $n$ optical modes. With a sequence of $O(n)$ tunable permutation modules, the circuit can be programmed to perform all the permutations required for the two tests (See Supplementary Information for details).
\par
In Stage (iii), the first layer of two-mode transformations $\{u_j(\theta_j)\}$ are all configured as two-mode Hadamard operations $H=\frac{1}{\sqrt{2n}}\begin{pmatrix}
1  & 1  \\
1 & -1 
\end{pmatrix}$ to interfere each of the $n/2$ pairs of the two optical modes $(2j-1,2j)$. The following mode routing rearranges the optical modes to enable possible further interferences required by the satisfiability test. This routing is realized by a high extinction-ratio polarizing beam-splitter (PBS). Two different types of configurations are adopted for the second layer of $\{u_j(\theta_j)\}$ depending on which of the satisfiability test and uniformity test is applied. If the uniformity test is chosen, all the transformations in this layer are set to $Z$ gates or identity operators $I=\frac{1}{\sqrt{n}}\begin{pmatrix}
1  & 0  \\
0 & 1 
\end{pmatrix}$ (without placing any operation on the two modes), which do not perform any interference. Therefore each mode corresponds to an outcome of the form $|i\rangle\pm |j\rangle$ for a certain matching $M$ in terms of the permutation. Arthur will reject the proof when the outcomes $\{|i\rangle+ |j\rangle,|i\rangle- |j\rangle\}$ both occur, that is, the two detectors (1,4),(2,5) or (3,6) labelled in Fig. \ref{fig:setup}c both click among the measurements on $K$ copies. For the satisfiability test, part of transformations in the last layer of $\{u_j(\theta_j)\}$ are set into two-mode Hadamard operations to further interfere two adjacent modes after the aforementioned mode routing. Finally, one of the output modes (the ``rejection mode'') for a group $(i,j,k,l)$ corresponds to the outcome $|i\rangle+|j\rangle+|k\rangle+|l\rangle$, thus Arthur can decide to reject or accept the proof based on whether the detector coupled to the rejection mode clicks (see Supplementary Information for details).
\par
The whole experimental set-up can form various Jamin–Lebedeff interferometers for different permutations and transformations. The beam displacers are strictly aligned and calibrated in order to maintain high interference visibilities for the interferometers when altering the permutations and transformations. The interference visibility for this type of interferometers is measured to be 99.4\%. Each of the six output modes of the circuit is coupled to a single-photon avalanche diode (Excelitas Technologies, SPCM-800-FC). Detection events are recorded by a time-correlated single-photon counting system (Swabian Instruments, Time Tagger Ultra) with a coincidence window of 4 ns. We register the measurement results of $5000 \times K$ photons for each test to provide the rejection probabilities shown in the figures.
\par
\textbf{Optical swap test.} Two single photons are injected into two proof encoding modules respectively to prepare the two quantum states $|\psi_1\rangle$ and $|\psi_2\rangle$, which yield the input field
\begin{eqnarray}
|\psi_{\text{in}}\rangle=&&|\psi_1\rangle |\psi_2\rangle \nonumber \\
=&&\left(\sum_{i=1}^n \alpha_{1,i} \hat{a}^\dagger_{1,i} {|0\rangle}_1\right)\left(\sum_{j=1}^n \alpha_{2,j} \hat{a}^\dagger_{2,j} {|0\rangle}_2\right) \nonumber \\
=&&\sum_{i,j}^n \alpha_{1,i}\alpha_{2,j} \hat{a}^\dagger_{1,i} \hat{a}^\dagger_{2,j} {|0\rangle}_1 {|0\rangle}_2.
\label{eq:hominput}
\end{eqnarray}
Here ${|0\rangle}_1$ and ${|0\rangle}_2$ represent the vacuum state for the two input sides. Then the two single-photon states interfere at the 50:50 NPBS for a multi-mode HOM interference. To observe the HOM interference, the fibre coupler labelled in Fig. \ref{fig:hom}a is moved by an electronically controlled translation stage (Thorlabs PT1-Z8) to change the relative delay between the wave packets of the two photons. The relationships between the creation operators for the input fields and output fields of the NPBS can be written as
\begin{eqnarray}
\hat{a}^\dagger_{1,i}=&&\frac{1}{\sqrt{2}}\left(\hat{a}^\dagger_{3,i}+\hat{a}^\dagger_{4,i}\right), \nonumber\\
\hat{a}^\dagger_{2,i}=&&\frac{1}{\sqrt{2}}\left(\hat{a}^\dagger_{3,i}-\hat{a}^\dagger_{4,i}\right). \label{eq:relation}
\end{eqnarray}
By substituting equation (\ref{eq:relation}) into equation (\ref{eq:hominput}), we obtain the output field 
\begin{eqnarray}
|\psi_{\text{out}}\rangle =&&\sum_{i,j}^n \frac{\alpha_{1,i}\alpha_{2,j}}{2} \left(\hat{a}^\dagger_{3,i}+\hat{a}^\dagger_{4,i}\right) \left(\hat{a}^\dagger_{3,j}-\hat{a}^\dagger_{4,j}\right) {|0\rangle}_3 {|0\rangle}_4 \nonumber \\
=&&\sum_{i} \frac{\alpha_{1,i}\alpha_{2,i}}{2}\left[\left(\hat{a}^\dagger_{3,i}\right)^2-\left(\hat{a}^\dagger_{4,i}\right)^2\right] {|0\rangle}_3 {|0\rangle}_4 \nonumber \\
&&+\sum_{i,j}^{i\neq j} \frac{\alpha_{1,i}\alpha_{2,j}}{2} \left(\hat{a}^\dagger_{3,i}\hat{a}^\dagger_{3,j}-\hat{a}^\dagger_{4,i}\hat{a}^\dagger_{4,j}\right) {|0\rangle}_3 {|0\rangle}_4 \nonumber \\
&&+\sum_{i,j}^{i\neq j} \frac{\alpha_{1,i}\alpha_{2,j}}{2} \left(\hat{a}^\dagger_{4,i}\hat{a}^\dagger_{3,j}-\hat{a}^\dagger_{3,i}\hat{a}^\dagger_{4,j}\right) {|0\rangle}_3 {|0\rangle}_4.
\end{eqnarray}
For indistinguishable photons, the resulting output state can be represented as
\begin{widetext}

\begin{eqnarray}
|\psi_{\text{out}}\rangle =&&\sum_{i} \frac{\alpha_{1,i}\alpha_{2,i}}{\sqrt{2}}\left( {|2_i\rangle}_3 {|0\rangle}_4-{|0\rangle}_3 {|2_i\rangle}_4\right) +\sum_{i,j}^{i<j} \frac{\alpha_{1,i}\alpha_{2,j}+\alpha_{1,j}\alpha_{2,i}}{2} \left({|1_i,1_j\rangle}_3 {|0\rangle}_4-{|0\rangle}_3 {|1_i,1_j\rangle}_4\right) \nonumber \\
&&+\sum_{i,j} \frac{\alpha_{1,i}\alpha_{2,j}-\alpha_{1,j}\alpha_{2,i}}{2} {|1_j\rangle}_3 {|1_i\rangle}_4.
\label{eq:psi_out}
\end{eqnarray}

\end{widetext}
Here ${|1_i,1_j\rangle}_3$ denotes the state with one photon in mode $i$ and another photon in mode $j$ for the output port 3. The right side of equation (\ref{eq:psi_out}) contains three terms, where the first two correspond to the one-side terms (two photons are in the same output port) and the last one corresponds to the two-side terms (one photon in the output port 3 and another photon in the output port 4). The probability of finding a ``two-side'' outcome is 
\begin{eqnarray}
p_r^{\text{swap}}=&&\frac{1}{4}\sum_{i,j} \left|\alpha_{1,i}\alpha_{2,j}-\alpha_{1,j}\alpha_{2,i}\right|^2 \nonumber \\
=&&\frac{1}{2}\sum_{i,j}|\alpha_{1,i}|^2 |\alpha_{2,j}|^2-\frac{1}{2}\sum_{i,j}\alpha_{1,i}^*\alpha_{1,j}\alpha_{2,i}\alpha_{2,j}^* \nonumber \\
=&&\frac{1}{2}\left(1-|\langle \psi_1|\psi_2\rangle|^2\right),
\end{eqnarray}
considering the overlap between the two states is $\langle \psi_1|\psi_2\rangle=\sum_{i}\alpha_{1,i}^*\alpha_{2,i}$. The probability $p_r^{\text{swap}}$ is consistent with the probability of finding a ``reject'' outcome in a swap test. In the experiment, each path mode of the output is attached to a SPAD, therefore the two polarization modes in the same path are detected by the same detector. This reduces the number of outcomes from $\dbinom{n}{2}$ to $\dbinom{n/2}{2}$. The coincidence channels $\{1,2\},\{1,3\},\{2,3\},\{4,5\},\{4,6\},\{5,6\}$ correspond to the ``accept'' outcome (here $\{i,j\}$ denotes a coincidence channel between detectors $i$ and $j$ as labelled in Fig. \ref{fig:hom}a). We also add photon number resolving detection by attaching a fiber beam-splitter (Thorlabs TN830R5F2) and an additional SPAD to two path modes. This scheme is capable of detecting more events on the ``accept'' outcome (see Supplementary Information for detailed results).

\par\noindent
\textbf{Acknowledgments}
\par\noindent
The authors thank N. Yu for helpful discussions. This work was supported by the National Key Research and Development Program of China (Grant Nos. 2019YFA0308704, 2017YFA0303703 and 2018YFB1003202), the National Natural Science Foundation of China (Grant Nos. 61972191, 11690032, 61490711, 11474159 and 91836303) and the Fundamental Research Funds for the Central Universities (Grant No. 020214380068). P. Y. acknowledges financial support by Anhui Initiative in Quantum Information Technologies (Grant No. AHY150100).


\clearpage
\begin{widetext}
\section{Supplementary Information}
\section{Experimental details}

\par
\textbf{Photon source. }The photon pairs generated by the SPDC source (denoted as signal and idler modes) were coupled into single mode fibers respectively. To characterize the second order coherence value $g^{(2)}(0)$ of heralded single photons, we detect photons with a single-photon avalanche diode
in the idler mode (denoted as mode $H$), of which a click heralds a single photon in the signal mode. The signal mode is split into two modes (labelled as $a$ and $b$) by applying a half-wave plate set in $22.5^{\circ}$ followed by a polarizing beam-splitter. The two modes are also detected by two single-photon avalanche diodes. We register the two-fold and three-fold coincidences between the detectors $a$, $b$ and $H$, then the second order coherence value can be calculated by
\begin{equation}
g^{(2)}(0)=\frac{C_{a,b,H}}{C_{a,H}C_{b,H}}N_H.
\end{equation}
Here $C_{a,b,H}$ denotes the three-fold coincidence rate between detectors $a,b$ and heralding, while $C_{a,H}$ ($C_{b,H}$) denotes the coincidence rate between detectors $a$ ($b$) and heralding. $N_H$ is the count rate of the heralding. We set the pump power of the source into different levels and measure the second order coherence values, as shown in Fig. \ref{fig:source}a. The measured values show an excellent agreement with the linear fitting.
\par
To observe the Hong-Ou-Mandel (HOM) interference between the two photons, we interfere the two photons by a non-polarizing beam-splitter (NPBS) and register the coincidence counts between the two detectors placed in the two output sides. The experimental results are shown in Fig. \ref{fig:source}b and the visibility of the HOM interference is $\mathcal{V}=0.969\pm 0.004$.

\begin{figure}[b!]
  \centering
  \includegraphics[width=\linewidth]{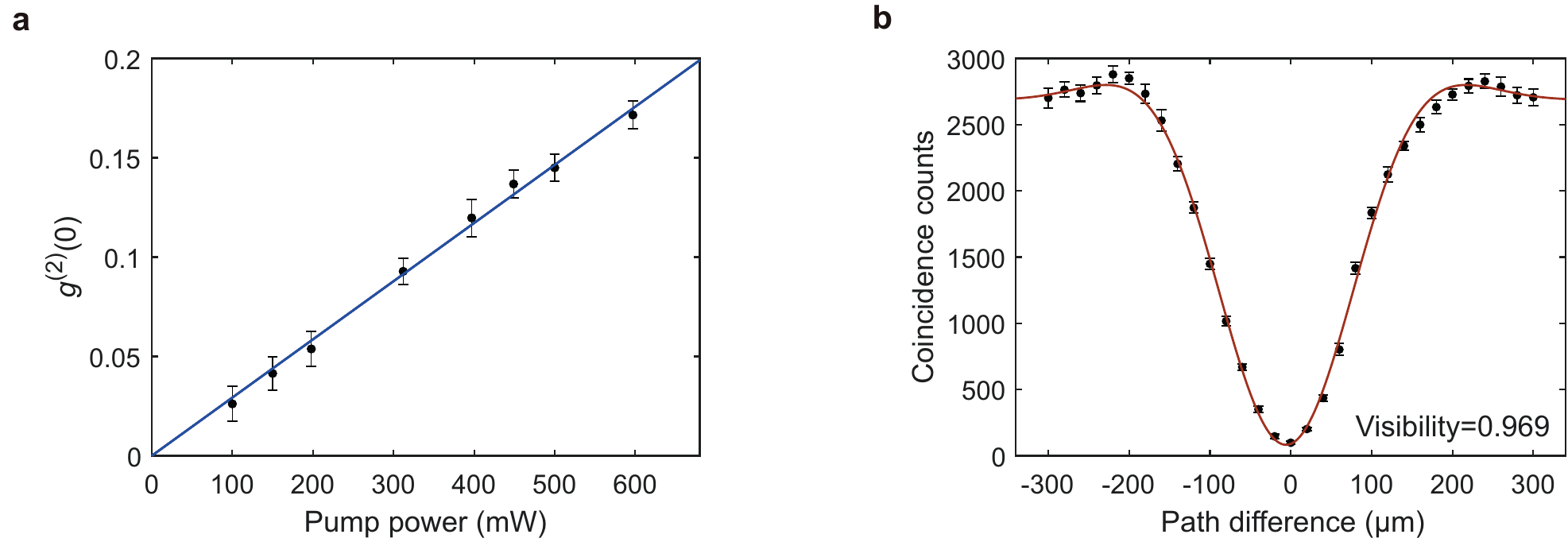}
\caption{\label{fig:source} Characterization of the photon source. (a) Experimental results of the $g^{(2)}$ measurements under different pump powers. The solid line (blue) is the linear curve fitting to the data. The errorbars are the standard uncertainties over 30 runs of the experiment. (b) The Hong-Ou-Mandel interference between the two photons. The solid line (red) is the curve fitting of the data to a Gaussian multiplied by sinc function. The errorbars are the standard uncertainties over 30 runs of the experiment.}
\end{figure}
\begin{figure}
  \centering
  \includegraphics[width=0.8\linewidth]{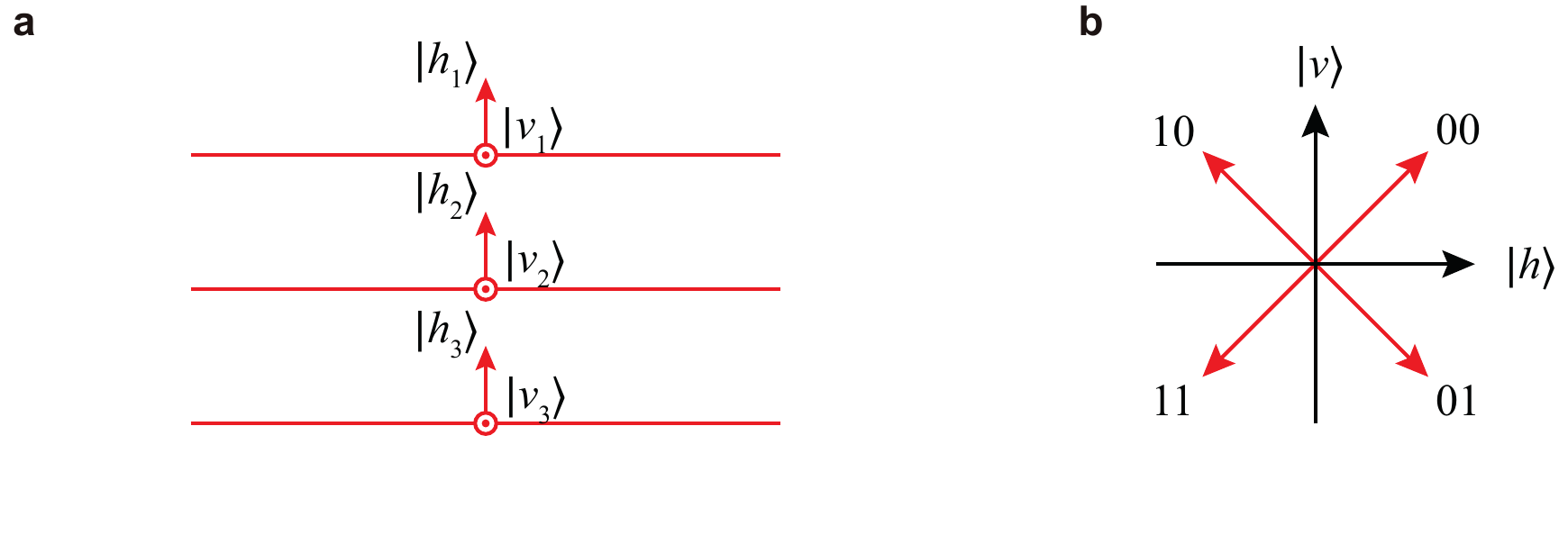}
\caption{\label{fig:encoding} (a) The mode mapping in the experiment. The horizontal and vertical polarizations in the three path modes $\{|h_1\rangle,|v_1\rangle,|h_2\rangle,|v_2\rangle,|h_3\rangle,|v_{3}\rangle\}$ correspond to the basis $\{|1\rangle,|2\rangle,|3\rangle,|4\rangle,|5\rangle,|6\rangle\}$ for the proof state. (b) The encoding in the polarization of a single path. The four polarization states correspond to the four possible values 00,01,10 and 11.}
\end{figure}

\par
\textbf{Proof encoding. }In the experiment, the proof is encoded in the polarization and path degrees of freedom of the photon. Each copy of the proof state is a superposition on 3 path modes $\times$ 2 polarization modes, which forms a 6-dimensional quantum state (see Fig. \ref{fig:encoding}). The optical modes $\{|h_1\rangle,|v_1\rangle,|h_2\rangle,|v_2\rangle,|h_3\rangle,|v_{3}\rangle\}$ correspond to the basis of the proof state $\{|1\rangle,|2\rangle,|3\rangle,|4\rangle,|5\rangle,|6\rangle\}$. Here $|h_1\rangle$ ($|v_1\rangle$) denotes the horizontal (vertical) polarization in path $1$. Figure \ref{fig:gates} shows the experimental implementation of the circuit operations on the six optical modes. The optics axis of each wave plate is pre-calibrated to guarantee accurate encoding and manipulation on the polarization in the following. The detailed procedure of proof encoding includes:
\par
(1) Prepare the photons in a coherent superposition of three path modes with equal amplitudes. The manipulation of path modes is achieved by combining half-wave plates to adjust the polarization and beam-displacers (BDs) that maps the polarization modes to path modes by moving the horizontally polarized photon with a 4 mm lateral displacement. In particular, we set the polarization of the single input path mode to $(\sqrt{2}|h_1\rangle+|v_1\rangle)/\sqrt{3}$ with a half-wave plate, then displace the horizontal polarization into another path mode. The polarization of the path mode is set into $(|h_1\rangle+|v_1\rangle)/\sqrt{2}$ and we displace the horizontal polarization with another BD to realize the superposition of three path modes.
\par
(2) Initialize the phase differences between the three path modes to zero. This is achieved by adjusting the tilt of each beam-displacer and confirmed by monitoring the interference between different path modes with a classical laser light.
\par
(3) After passing the splitting module, the single photonic state undergoes a combination of three unitary transformations $u_j(\theta_j)$. The unitary transformations are implemented by three half-wave plates with electronically-controlled rotation stages (Newport PR50PP). Each wave plate is configured into one of the four angles ($-67.5^{\circ},-22.5^{\circ},22.5^{\circ},67.5^{\circ}$ or $-22.5^{\circ},22.5^{\circ},67.5^{\circ},112.5^{\circ}$) to realize one of the four sub-operations
\begin{equation}
\frac{1}{\sqrt{2n}}\begin{pmatrix}
-1  & -1  \\
-1 & 1 
\end{pmatrix},\frac{1}{\sqrt{2n}}\begin{pmatrix}
1  & -1  \\
-1 & -1 
\end{pmatrix},\frac{1}{\sqrt{2n}}\begin{pmatrix}
1  & 1  \\
1 & -1 
\end{pmatrix},\frac{1}{\sqrt{2n}}\begin{pmatrix}
-1  & 1  \\
1 & 1 
\end{pmatrix}. \nonumber
\end{equation}
Each sub-operation prepares a certain polarization corresponding to the encoding of two variables (see Fig. \ref{fig:encoding}b). For example, the polarization states $\{(|h_1\rangle+|v_1\rangle)/\sqrt{2},(|h_1\rangle-|v_1\rangle)/\sqrt{2},(-|h_1\rangle+|v_1\rangle)/\sqrt{2},(-|h_1\rangle-|v_1\rangle)/\sqrt{2}\}$ correspond to the encoding of the assignments 00,01,10,11 into the subspace of the proof state respectively. Consequently, the three reconfigurable wave plates enable the encoding of all the 64 possible assignments into the proof state (i.e., the 64 proper states). Note the proof state can also be prepared into improper states by setting the wave plates in the proof encoding stage into other angles.

\begin{figure}
  \centering
  \includegraphics[width=0.8\linewidth]{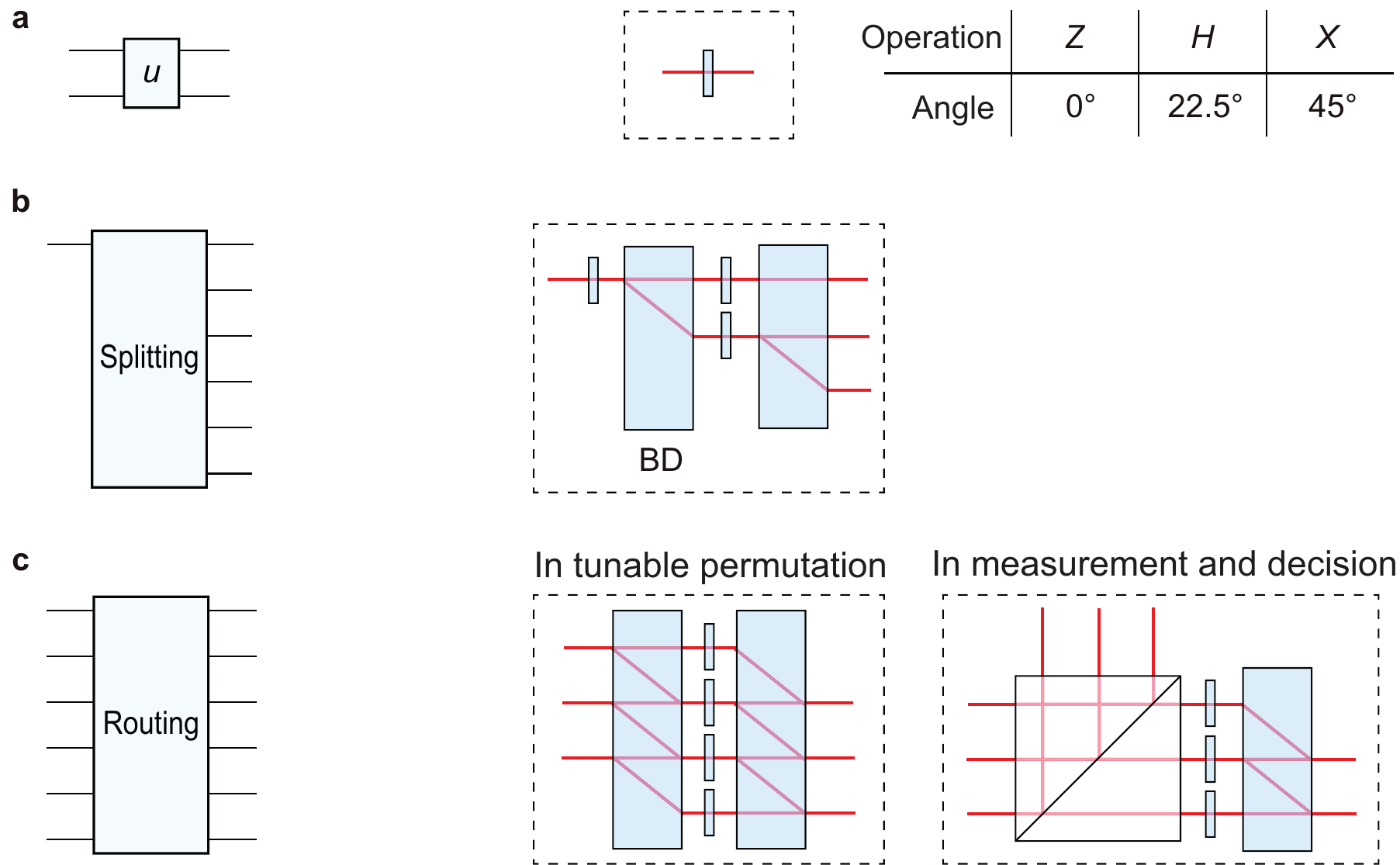}
\caption{\label{fig:gates} The correspondence between the circuit operations and the experimental setups. (a-c) The experimental implementations of the unitary transformation, the mode splitting and the mode routing.}
\end{figure}

\begin{figure}
  \centering
  \includegraphics[width=0.8\linewidth]{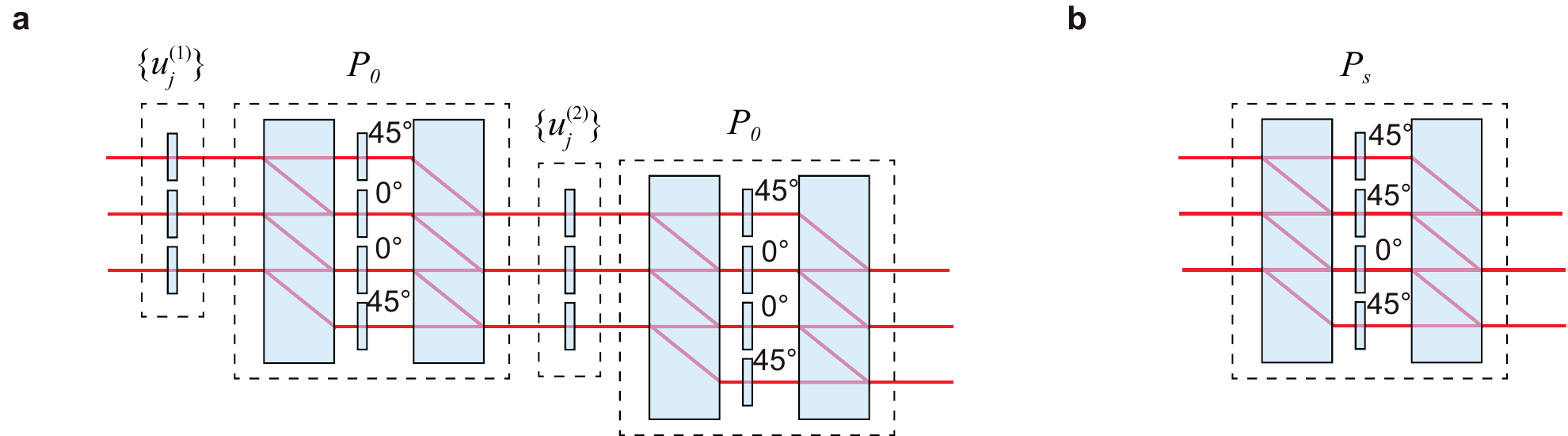}
\caption{\label{fig:routing} The detailed settings of the mode routing in the tunable permutations stage. (a) The stage consists of two modules, each with a layer of unitary transformations $\{u_j(\theta_j)\}$ and a mode routing. For most of the cases in our experiments, the two mode routing modules are set into $P_0$. (b) For the case of verifying the clause $(1,2,5,6)$, the first mode permutation is tuned to $P_s$ by changing the operations of one of the wave plates.}
\end{figure}

\begin{figure}
  \centering
  \includegraphics[width=\linewidth]{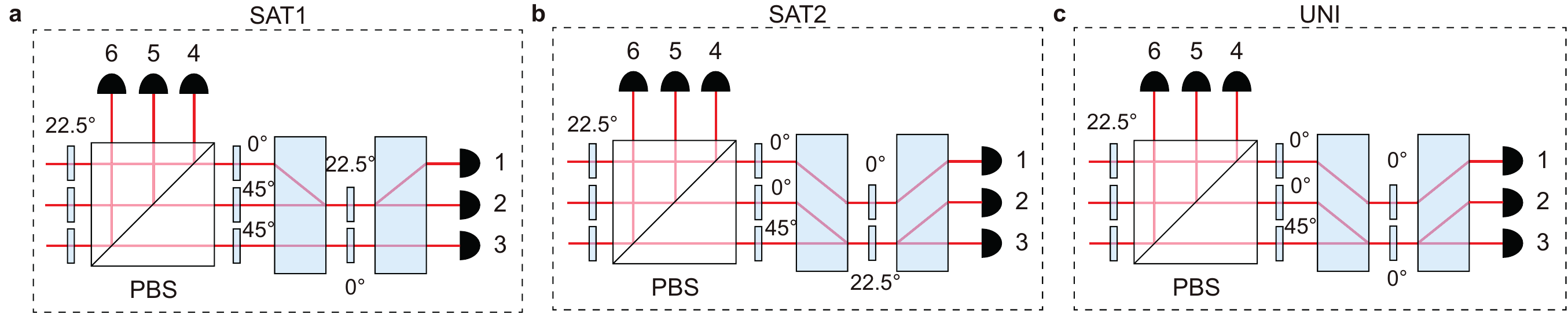}
\caption{\label{fig:mea_set} The detailed settings of the measurement and decision stage. (a) The settings of the wave plates to realize the measurement ``SAT1'', which performs projection on the modes (1,2,3,4) after the permutation. (b) The settings of the wave plates to realize the measurement ``SAT2'', which performs projection on the modes (3,4,5,6) after the permutation. (c) The settings of the wave plates to realize the measurement ``UNI'', which performs interference between the pairs of optical modes after the permutation.}
\end{figure}

\par
\textbf{Arthur's computation machine. }For each copy of the proof state, Arthur firstly performs the tunable permutations on the optical modes in Stage (ii). As explained in the main text, each module of tunable permutation is composed of a layer of unitary transformations $\{u_j(\theta_j)\}$ and a mode routing $P$. In our experiment, the unitary transformations can be set into different operations such as 
\begin{equation}
Z=\frac{1}{\sqrt{n}}\begin{pmatrix}
1  & 0  \\
0 & -1 
\end{pmatrix},
H=\frac{1}{\sqrt{2n}}\begin{pmatrix}
1  & 1  \\
1 & -1 
\end{pmatrix},
X=\frac{1}{\sqrt{n}}\begin{pmatrix}
0 & 1  \\
1 & 0 
\end{pmatrix} \nonumber
\end{equation}
by different settings of the wave plates (see Fig. \ref{fig:gates}a). Regarding the mode routing (see Fig. \ref{fig:gates}c, the first box), we can set the wave plates in each module to realize one of the two permutation matrices 
\begin{equation}
P_0=\begin{pmatrix}
 0& 1& 0& 0& 0& 0 \\
 0& 0& 0& 1& 0& 0 \\
 1& 0& 0& 0& 0& 0 \\
 0& 0& 0& 0& 0& 1 \\
 0& 0& 1& 0& 0& 0 \\
 0& 0& 0& 0& 1& 0
 \end{pmatrix},
 P_s=\begin{pmatrix}
 0& 1& 0& 0& 0& 0 \\
 1& 0& 0& 0& 0& 0 \\
 0& 0& 0& 1& 0& 0 \\
 0& 0& 0& 0& 0& 1 \\
 0& 0& 1& 0& 0& 0 \\
 0& 0& 0& 0& 1& 0
 \end{pmatrix}. \nonumber
\end{equation}
The configurations of the wave plates for implementing the two permutation matrices are depicted in Fig. \ref{fig:routing}. Note the operations of the aforementioned two routing modules have negative values $-1$ in some of the elements in actual implementations, due to the fact that a Pauli-$Z$ sub-operation (a half-wave plate aligned in $0^{\circ}$) adds a relative $\pi$ phase in one of the two optical modes. Yet in the following design of configurations we have take the phases added by all the wave plates into consideration to eliminate the effect, therefore the modules equivalently realize the permutations $P_0$ and $P_1$. In our experiments, Stage (ii) utilizes two modules of the tunable permutations, as shown in Fig. \ref{fig:routing}a. The two layers of unitary transformations are denoted as $\{u^{(1)}_j(\theta_j^{(1)})\}$ and $\{u^{(2)}_j(\theta_j^{(2)})\}$ respectively and the overall transformations are thus $U^{(1)}=\bigoplus_{j=1}^{n/2}u^{(1)}_j(\theta_j^{(1)})$ and $U^{(2)}=\bigoplus_{j=1}^{n/2}u^{(2)}_j(\theta_j^{(2)})$. The two mode routing modules are normally set into $P_0$ for the cases of verifying 14 clauses and all the 15 matchings, whereas the first mode routing is set into $P_s$ for the case of verifying the clause $(1,2,5,6)$. The detailed configurations of waveplates to implement the 15 projections and the 15 matchings are given in Table \ref{tab:setting_sat} and Table \ref{tab:setting_uni} respectively.

\par
In Stage (iii), Arthur resorts to one of the three types of measurements depending on which test is chosen and which permutation is performed. The settings of the three types of measurements are depicted in Fig. \ref{fig:mea_set}. For all the three types of measurements, the state firstly undergoes a layer of unitary transformations set into $H$ operations and a mode routing (Fig. \ref{fig:gates}c, the second box). The permutation matrix for this mode routing (in the measurement and decision part) can be described as
\begin{equation}
P_m=\begin{pmatrix}
 1& 0& 0& 0& 0& 0 \\
 0& 0& 1& 0& 0& 0 \\
 0& 0& 0& 0& 1& 0 \\
 0& 1& 0& 0& 0& 0 \\
 0& 0& 0& 1& 0& 0 \\
 0& 0& 0& 0& 0& 1
 \end{pmatrix}. \nonumber
\end{equation}
For the satisfiability test, Arthur then performs the measurement ``SAT1'' or ``SAT2'', depending on which four variables are verified (upper or lower). For the uniformity test, Arthur performs the measurement ``UNI'' to interfere the pairs of the optical modes.
\begin{table}
 \centering \caption{\label{tab:setting_sat}Detailed configurations of Arthur's setup for the satisfiability test. The rejection mode denotes the optical mode corresponding to the projection on $|c\rangle$, therefore a click on the mode leads to the ``reject'' decision. Note for the clause $(1,2,5,6)$, the first mode routing in Stage (ii) is set into $P_s$, whereas for the other 14 clauses the mode routing is set into $P_0$.}
\begin{ruledtabular}
\begin{tabular}{cccccccccc}
   & Clause    & \multicolumn{6}{c}{Transformations} & Measurement & Rejection mode \\ 
   &           & $u^{(1)}_1$ & $u^{(1)}_2$ & $u^{(1)}_3$ & $u^{(2)}_1$ & $u^{(2)}_2$ & $u^{(2)}_3$ & &   \\ \hline
1  & (1,2,3,4) & X    & X    & X   & X   & X   & X   & SAT1        & 2                    \\
2  & (1,2,3,5) & Z    & Z    & Z   & Z   & Z   & Z   & SAT2        & 2                    \\
3  & (1,2,3,6) & Z    & Z    & X   & Z   & Z   & Z   & SAT2        & 2                    \\
4  & (1,2,4,5) & X    & X    & Z   & Z   & Z   & X   & SAT2        & 3                    \\
5  & (1,2,4,6) & Z    & X    & X   & Z   & Z   & Z   & SAT2        & 2                    \\
6  & (1,2,5,6) & X    & X    & X   & X   & Z   & Z   & SAT1        & 2                    \\
7  & (1,3,4,5) & Z    & X    & Z   & X   & Z   & X   & SAT2        & 3                    \\
8  & (1,3,4,6) & X    & X    & Z   & X   & Z   & X   & SAT1        & 2                    \\
9  & (1,3,5,6) & X    & Z    & X   & Z   & X   & X   & SAT2        & 3                    \\
10 & (1,4,5,6) & X    & X    & X   & Z   & X   & X   & SAT2        & 3                    \\
11 & (2,3,4,5) & X    & X    & Z   & X   & Z   & X   & SAT2        & 3                    \\
12 & (2,3,4,6) & X    & X    & X   & X   & Z   & X   & SAT2        & 3                    \\
13 & (2,3,5,6) & Z    & X    & X   & Z   & Z   & Z   & SAT1        & 2                    \\
14 & (2,4,5,6) & Z    & Z    & Z   & Z   & Z   & Z   & SAT1        & 1                    \\
15 & (3,4,5,6) & X    & X    & X   & X   & X   & X   & SAT2        & 3                    \\
\end{tabular}
\end{ruledtabular}
\end{table}

\begin{table}
 \centering \caption{\label{tab:setting_uni}Detailed configurations of Arthur's setup for the uniformity test. For all the 15 matchings the two mode routing modules in Stage (ii) are both set into $P_0$.}
\begin{ruledtabular}
\begin{tabular}{ccccccccc}
   & Matching          & \multicolumn{6}{c}{Transformations} & Measurement\\
   &                   & $u^{(1)}_1$ & $u^{(1)}_2$ & $u^{(1)}_3$ & $u^{(2)}_1$ & $u^{(2)}_2$ & $u^{(2)}_3$ &     \\ \hline
1  & (1,2),(3,4),(5,6) & X    & X    & X   & X   & X   & X   & UNI \\
2  & (1,2),(3,5),(4,6) & X    & Z    & X   & X   & X   & Z   & UNI \\
3  & (1,2),(3,6),(4,5) & X    & Z    & Z   & X   & X   & Z   & UNI \\
4  & (1,3),(2,4),(5,6) & X    & Z    & X   & Z   & X   & X   & UNI \\
5  & (1,3),(2,5),(4,6) & Z    & Z    & Z   & Z   & Z   & Z   & UNI \\
6  & (1,3),(2,6),(4,5) & Z    & Z    & X   & Z   & Z   & Z   & UNI \\
7  & (1,4),(2,3),(5,6) & X    & X    & X   & Z   & X   & X   & UNI \\
8  & (1,4),(2,5),(3,6) & X    & X    & Z   & Z   & Z   & X   & UNI \\
9  & (1,4),(2,6),(3,5) & Z    & X    & X   & Z   & Z   & Z   & UNI \\
10 & (1,5),(2,3),(4,6) & X    & Z    & Z   & Z   & Z   & Z   & UNI \\
11 & (1,5),(2,4),(3,6) & X    & X    & Z   & Z   & Z   & Z   & UNI \\
12 & (1,5),(2,6),(3,4) & X    & X    & X   & X   & Z   & X   & UNI \\
13 & (1,6),(2,3),(4,5) & X    & Z    & Z   & X   & Z   & Z   & UNI \\
14 & (1,6),(2,4),(3,5) & X    & X    & Z   & X   & Z   & Z   & UNI \\
15 & (1,6),(2,5),(3,4) & X    & X    & Z   & X   & Z   & X   & UNI \\
\end{tabular}
\end{ruledtabular}
\end{table}

\par
\textbf{Experimental imperfections. } In the satisfiability test, each combination of a proof state and a verified clause corresponds to a rejection probability $p_c$. To characterize the experimental errors, we calculate the average statistical fidelity 
\begin{equation}
\mathcal{F}_c=\left(\sqrt{p_c^\text{the} p_c^\text{exp}}+\sqrt{(1-p_c^\text{the}) (1-p_c^\text{exp}})\right)^2
\end{equation}
between the theoretical and experimental projection probabilities ($p_c^\text{the}$ and $p_c^\text{exp}$).  The limited interference visibility and the phase fluctuations, together with the systematic errors in the operations, are responsible for the deviations and result in a non-zero projection probability for the satisfying proofs. The alignments of BDs are pre-calibrated to achieve interference visibilities exceeding 99\% and the path differences within the interferometers are tuned to zero. Regarding the phase fluctuations, the compact interferometers implemented by the beam displacers (BDs) are stable against environmental perturbations. The paths split by BDs are parallel to each other and the distance between the adjacent paths is a 4 mm lateral displacement, which ensures that the paths undergo nearly the same phase fluctuations. Therefore the phase difference between different paths is passively stabilized. In addition, we built an optical enclosure to shield the experimental set-up from environmental variations to further suppress the phase fluctuations. As a result, the whole set-up can remain stable in a time scale of 3 hours. The systematic errors mainly stem from the misalignments of the wave plates and the beam displacers. In our experiments, the optics axis of each wave plate is calibrated independently with a precision of $0.1^{\circ}$. The repetition errors on the angles of electronically-controlled wave plates (Newport PR50PP) are typically $0.025^{\circ}$. In addition, the unbalanced detection efficiencies for the optical modes may cause deviations of the outcome probabilities. In our experiment we adjust the coupling efficiencies of the optical modes to balance the overall detection efficiencies. It is noteworthy that the scheme only requires at most two layers of cascaded interferometers. Therefore, we expect that the scheme remains a high fidelity even scaled to large size.
\par
For the optical swap test experiments, the main errors include the limited photon indistinguishability from the source part, the non-ideal splitting ratio of the NPBS and the unbalanced detection efficiencies for different detectors. The path differences between the three paths for the two input sides are calibrated by interferometers with classical light.
\par
When scaling the scheme up to higher $n$, the limited multi-mode interference visibility and HOM visibility would reduce the completeness-soundness gap with the increase of the number of copies $K$, due to the fact that the completeness is not perfect in practical realizations. However, the QMA(2) protocol allows amplification of success probability by repeating the original protocol, which we demonstrate in Sec. \ref{amplify}.
\par
Another factor in practical experiments is the photon loss. The copies of proof states may not be detected by Arthur due to the photon loss in his computation machine. The decision of each test in our experiment is based on the detection of $K$ photons. On average, Merlins need to send $O(K/\eta)$ photons if the overall photon efficiency of Arthur's machine is $\eta$. In our experiment, the photon efficiency, accounting for the transmission of the linear optical circuit ($\sim 80\%$), the fiber coupling efficiency ($\sim 90\%$) and the detection efficiency of the SPADs ($\sim 60\%$), is about 43\%.

\section{Experimental results}
\begin{figure*}
  \centering
  \includegraphics[width=0.95\linewidth]{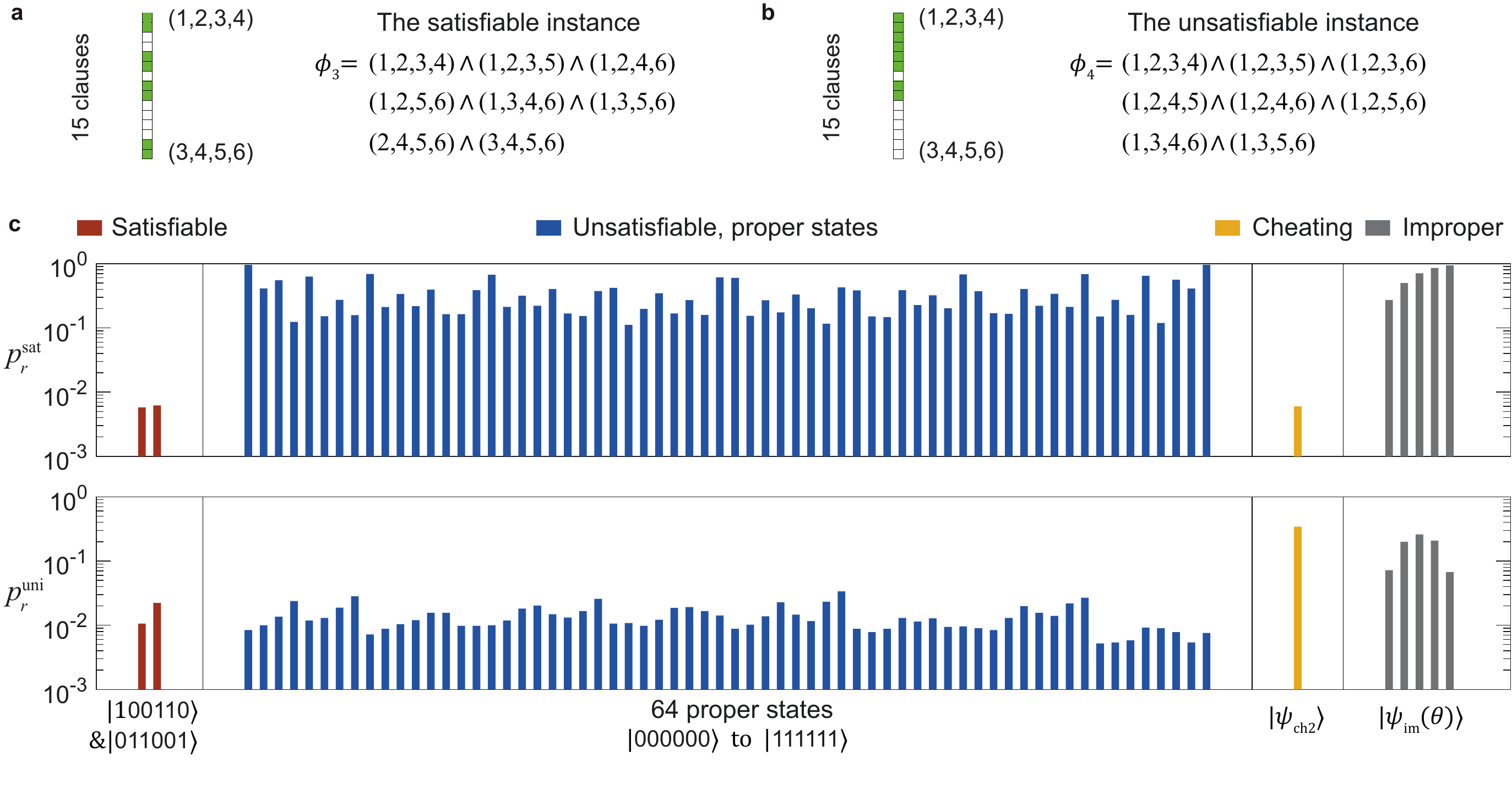}
\caption{\label{fig:example2} \textbf{Experimental verification of example instances.} \textbf{a}, The satisfiable instance $\phi_3$. \textbf{b}, The unsatisfiable instance $\phi_4$. The shaded squares (green) illustrate which 8 of the 15 clauses are chosen in the instance. \textbf{c}, The rejection probabilities of the satisfiability test ($p_r^\text{sat}$, top) and uniformity test ($p_r^\text{uni}$, down) for different proof states. For the satisfiable instance $\phi_3$, Merlins will send states encoding the correct solution thus we show the results for the two satisfying proof states (red bars). For the unsatisfiable instance $\phi_4$, we test different cases consisting of sending the 64 proper states (blue bars), a deliberate cheating proof state $|\psi_\text{ch2}\rangle$ in order to pass the satisfiability test (yellow bars), improper states $|\psi_{\text{im}}(\theta)\rangle$ (grey bars). The number of copies $K=3$ is adopted in the verification.} 
\end{figure*}

\begin{figure}
  \centering
  \includegraphics[width=0.9\linewidth]{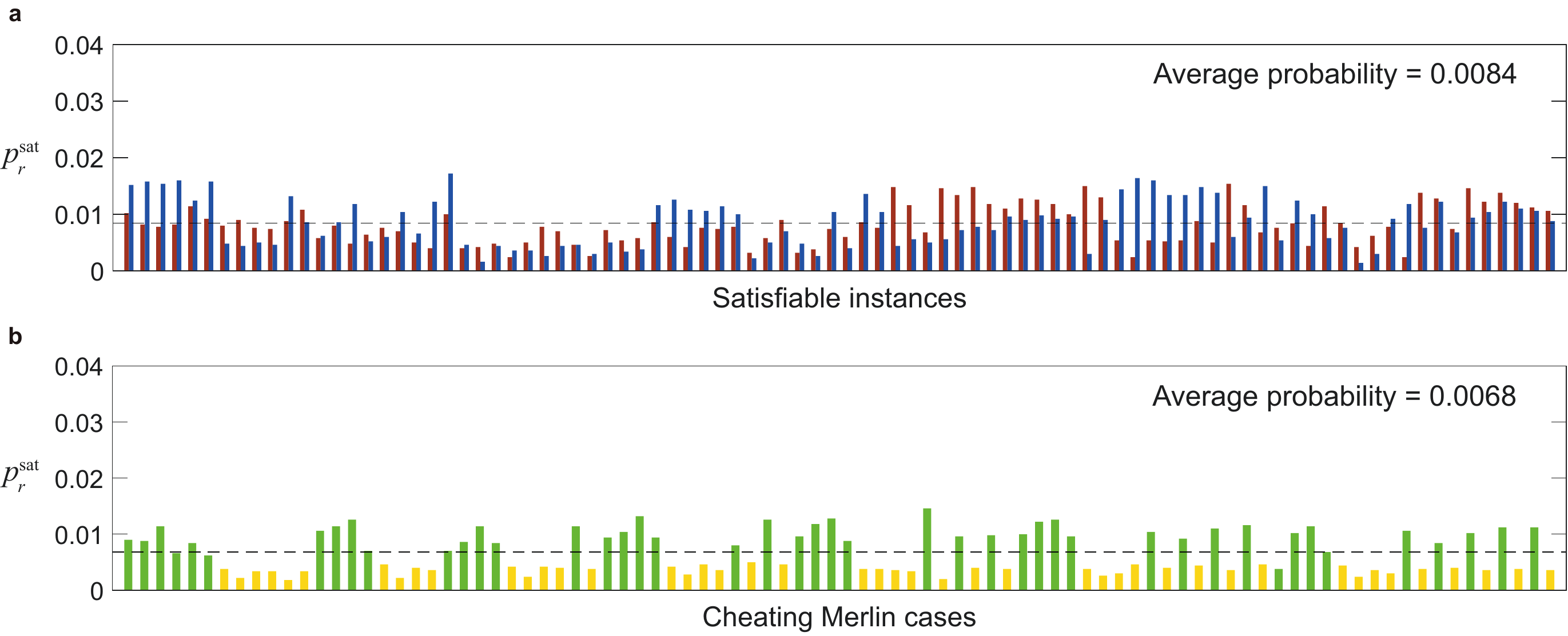}
\caption{\label{fig:detail} (a) The rejection probabilities of the satisfiability test for all the 90 satisfiable instances. For each instance, there are two satisfying assignments (proof states), which are denoted as red bars and blue bars respectively. (b) The rejection probabilities of the satisfiability test for the cheating cases. In each case, Merlins send $|\psi_{\text{ch1}}\rangle$ (yellow bars) or $|\psi_{\text{ch2}}\rangle$ (green bars) as proof states.}
\end{figure}

\par
\textbf{More examples and detailed results.} As a supplement of the experimental results of quantum verification shown in Fig. 3 in the main text, we demonstrate the quantum verification of other example instances, including an unsatisfiable instance where cheating Merlins send copies of $|\psi_{\text{ch2}}\rangle$, as shown in Fig. \ref{fig:example2}. To examine the completeness of the protocol, we test the performance of the algorithm in verifying all the 90 satisfiable instances. Figure \ref{fig:detail}a shows the detailed results when verifying the satisfiable instances. For all the 90 instances Arthur has high probabilities to accept, which confirm the nearly perfect completeness of the protocol. In addition, we also give the results for the 90 cheating cases where Merlins send the states $|\psi_{\text{ch1}}\rangle$ or $|\psi_{\text{ch2}}\rangle$ in Fig. \ref{fig:detail}b. The rejection probabilities for these cheating cases are in the same order of magnitude as the probabilities for verifying satisfiable instances.
\begin{figure}
  \centering
  \includegraphics[width=\linewidth]{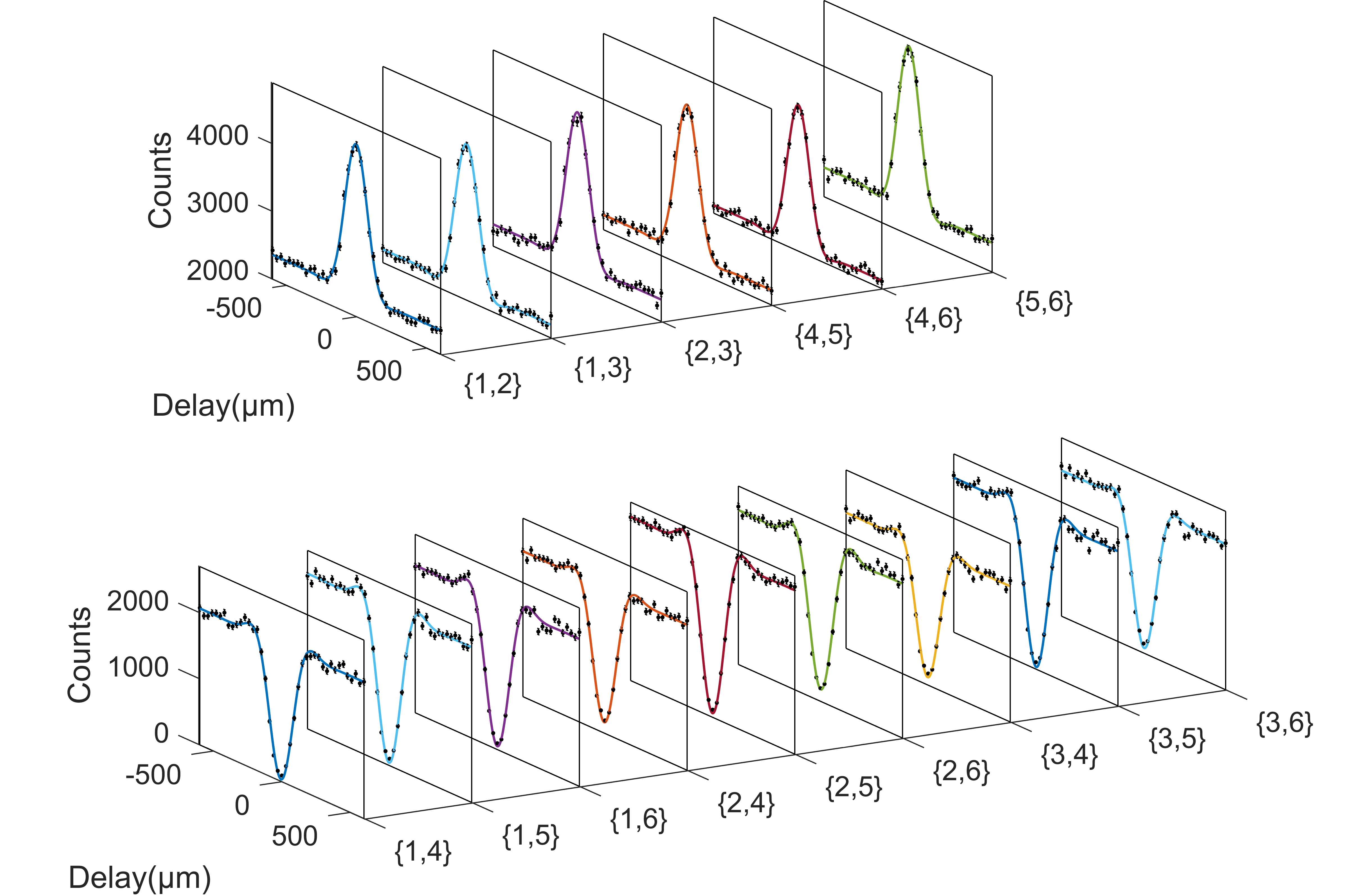}
\caption{\label{fig:hom} Observation of Hong-Ou-Mandel (HOM) interference for the 15 coincidence channels. There are 6 one-side channels corresponding to the ``accept'' output (the upper panel) and 9 two-side channels corresponding to the ``reject'' output (the lower panel). Solid lines are curve fittings of the data (black dots) to a Gaussian multiplied by sinc function. Error bars are uncertainties assuming Poisson count statistics.}
\end{figure}
\begin{figure}
  \centering
  \includegraphics[width=0.4\linewidth]{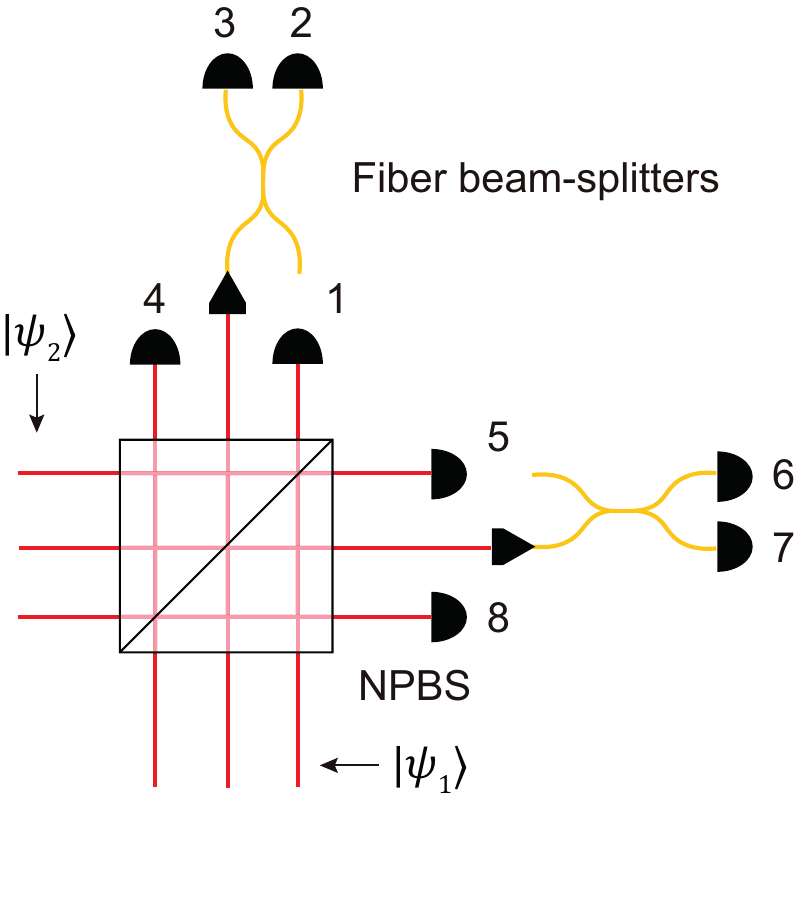}
\caption{\label{fig:8det} Optical swap test with additional detectors. The HOM interference scheme is as described in the main text (Fig. 4a). Two 50:50 fiber beam-splitters are attached to the middle optical path in the two output sides respectively. Additional SPADs are also coupled to the outputs of the fiber beam-splitters. The two-fold coincidences for the clicks of the 8 detectors are registered.}
\end{figure}
\begin{figure}
  \centering
  \includegraphics[width=\linewidth]{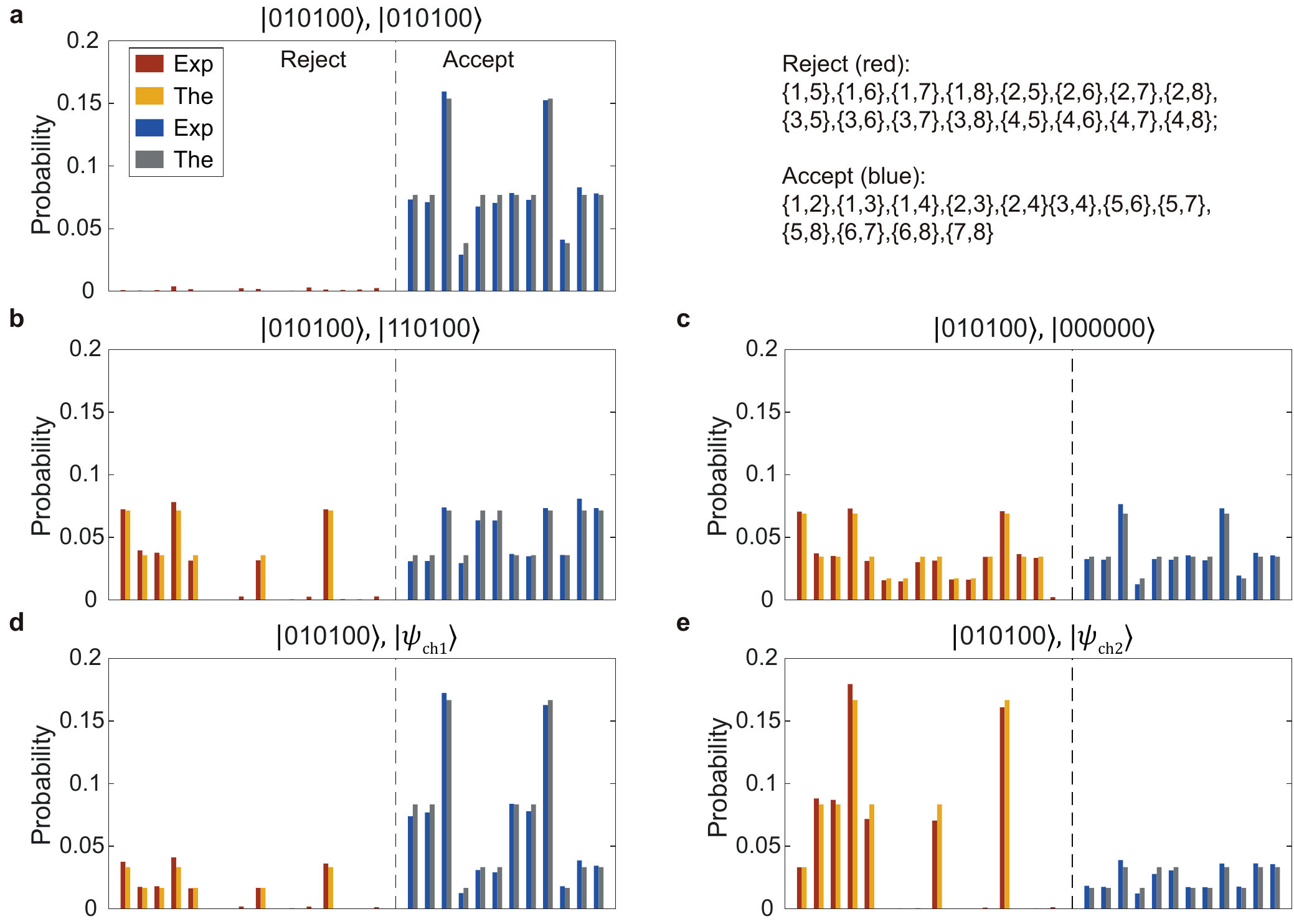}
\caption{\label{fig:8det_results} The results of optical swap test with additional detectors for typical cases: (a) the two states are proper and identical; (b,c) the two states are proper but not identical, (d,e) one of the state is proper and the another is improper. Each panel shows the experimental (red and blue bars) and theoretical (yellow and grey bars) outcome probabilities on the 28 coincidence channels.}
\end{figure}
\par
\textbf{Optical swap test. } The results of the Hong-Ou-Mandel (HOM) interference shown in the main text are based on the detection events of the 15 two-fold coincidence channels. Here we observe the HOM interference for all the 15 outcomes, as illustrated in Fig. \ref{fig:hom}. The interferences for the 6 ``accept'' channels manifest peaks whereas the interferences for the 9 ``reject'' channels manifest dips. In visualizing the results of the HOM interference in Fig. 4b in the main text, a factor of $3/2$ is applied to the probability of the ``accept'' outcome to compensate the events that two photons trigger the same detector. To demonstrate the detection of events that two photons are in the same path, we also implement the optical swap test with additional detectors, as depicted in Fig. \ref{fig:8det}. The additional detectors add photon-number resolution to two of the optical paths. Under this detection scheme, we perform the optical swap tests on states that are the same as in the main text. For the case that the two states are the same, higher acceptance probability is observed compared the results with 6 detectors.

\section{Amplification of the success probability}\label{amplify}
\begin{figure}
  \centering
  \includegraphics[width=0.5\linewidth]{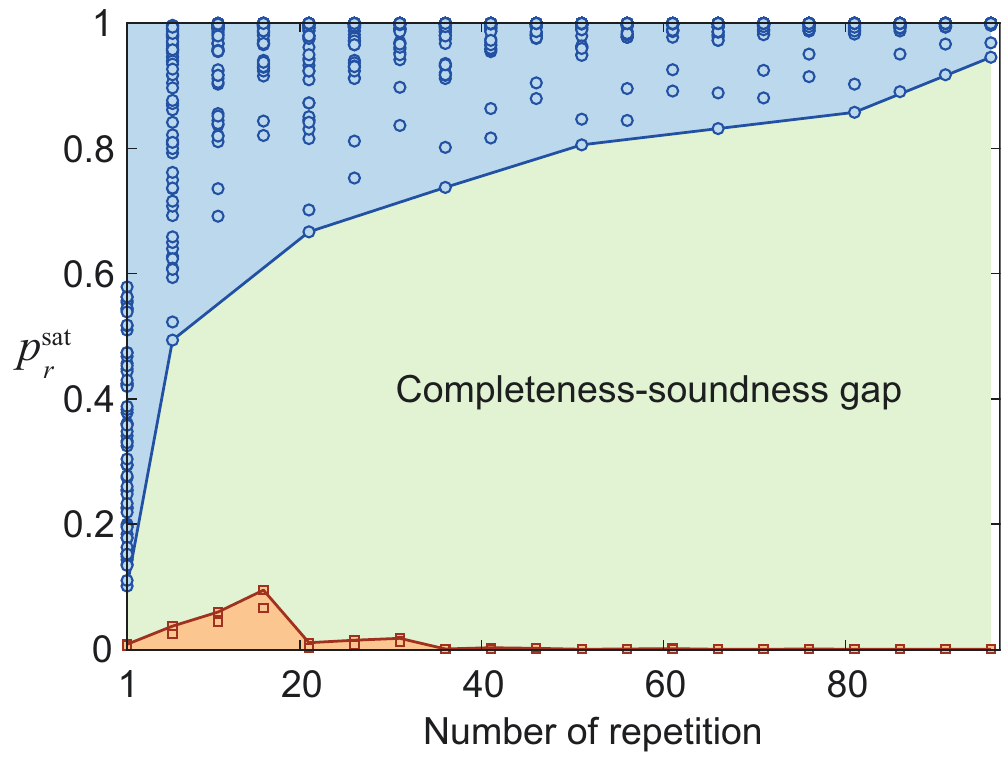}
\caption{\label{fig:am} The amplification of the completeness-soundness gap. Each marker represents the rejection probability of a proof in verifying the example satisfiable instance $\phi_1$ (circle, blue) and the example unsatisfiable instance $\phi_2$ (square, red) in the main text. The success probability of the verification (green area) increases with the increase of the number of repetition.}
\end{figure}
\par
For practical use of verification algorithms, a problem of particular interest is the ability to amplify the success probability of Arthur's decision. It has been conjectured and proven that any QMA(2) protocol can be amplified to exponentially small error~\cite{v005a001,Harrow:2013:TPS:2432622.2432625}. Here we resort to an amplification protocol proposed in Ref.~\cite{v005a001} to demonstrate the amplification of the success probability. The main idea is to repeat the original verification protocol a certain number of times and then output an answer based on specific criteria. For example, Merlins and Arthur firstly run the verification protocol $T$ times, then Arthur accepts if at least $(c+s)T/2$ runs of the verification algorithm output ``accept'' and rejects otherwise. Here $c$ and $s$ represent completeness and soundness of the original protocol respectively. As an example, we perform the amplification protocol in verifying the satisfiability of instances $\phi_1$ and $\phi_2$ used in the main text. The success probability of the verification increases with the increase of the number of repetition, as shown in Fig. \ref{fig:am}.

\end{widetext}

\end{document}